\documentclass[12pt,reqno]{amsart}

\usepackage[margin=1in,dvips,a4paper]{geometry}
\usepackage[margin=1em,font=small,justification=centerlast]{caption}
\usepackage[dvips,svgnames]{xcolor}

\usepackage{pstricks,pst-node}

\psset{unit=1bp,linewidth=.5bp,arrowsize=3bp 2}

\def\arrowLine(#1,#2)(#3,#4){%
  \pcline(#1,#2)(#3,#4)%
  \lput{:U}{
    \pspicture(0,0)(0,0)
      \psline[arrows=->](1.2,0)(1.3,0)
    \endpspicture
  }
}

\def \bbC{\mathbb C}
\def \rmd{\mathrm d}
\def \rme{\mathrm e}
\def \rmi{\mathrm i}
\def \calR{\mathcal R}

\newtheorem{theorem}{Theorem}
\newtheorem{proposition}{Proposition}

\theoremstyle{definition}
\newtheorem*{example}{Example}

\begin{document}

\title[Polynomial solutions of $q$KZ equation]{Polynomial solutions of $q$KZ equation and\\
ground state of XXZ spin chain at $\Delta = -1/2$}
\author{A.~V.~Razumov}
\address{A.~V.~Razumov, Division of Theoretical Physics\\
Institute for High Energy Physics\\
142281, Protvino\\
Moscow region, Russia}
\author{Yu.~G.~Stroganov}
\address{Yu.~G.~Stroganov, Division of Theoretical Physics\\
Institute for High Energy Physics\\
142281, Protvino\\
Moscow region, Russia}
\author{P.~Zinn-Justin}
\address{P.~Zinn-Justin, Laboratoire de Physique Th\'eorique et Mod\`eles Statistiques\\ Universit\'e Paris-Sud\\
F-91405 Orsay, France}

\date{\today}

\begin{abstract}
Integral formulae for polynomial solutions of the quantum Knizhnik--Zamolodchikov equations associated with the $R$-matrix of the six-vertex model are considered. It is proved that when the deformation parameter $q$ is equal to $\rme^{\pm 2 \pi \rmi/3}$ and the number of vertical lines of the lattice is odd, the solution under consideration is an eigenvector of the inhomogeneous transfer matrix of the six-vertex model. In the homogeneous limit it is a ground state eigenvector of the antiferromagnetic XXZ spin chain with the anisotropy parameter $\Delta$ equal to $-1/2$ and odd number of sites. The obtained integral representations for the components of this eigenvector allow to prove some conjectures on its properties formulated earlier. A new statement relating the ground state components of XXZ spin chains and Temperley--Lieb loop models is formulated and proved.
\end{abstract}

\maketitle

\section{Introduction}

The recent years have witnessed an explosion of conjectures concerning the ground state of the antiferromagnetic XXZ spin chain with the anisotropy parameter $\Delta$ equal to $-1/2$ \cite{Str01, RazStr01, BatdeGNie01, RazStr01b} and of a closely related Temperley--Lieb loop model \cite{BatdeGNie01, RazStr04, PeaRitdeG01, RazStr05, PeaRitdeGNie02, DiF04a, deGRit04, DiF04b}. They provide an interesting connection to the world of combinatorics, and in particular to the realm of alternating sign matrices \cite{Bres}. Some progress has been made towards understanding these conjectures by use of the connection of these one-dimensional quantum mechanical models with two-dimensional integrable models of statistical mechanics, the six-vertex model and the dense O$(1)$ loop model respectively. Here instead of eigenvectors of the Hamiltonians one studies eigenvectors of the transfer matrices. An important point of this approach, initiated by Di~Francesco and Zinn-Justin in the paper \cite{DiFZin05a}, is the transition to inhomogeneous versions of the associated models of statistical mechanics. It eventually led to the idea that one should replace the original eigenvector equation defining the ground state with the quantum Knizhnik--Zamolodchikov ($q$KZ) equation \cite{DiFZin05b}. The $q$KZ equation contains a free parameter $q$ that is related to the anisotropy parameter $\Delta$ of the XXZ spin chain by $\Delta=(q+q^{-1})/2$, so that one should set $q=\rme^{\pm 2\pi\rmi/3}$ in the end.

Much is known about solutions of the $q$KZ equation. As in the closely related work~\cite{DiFZin07}, we shall here write integral formulae for its solutions. The focus of the present paper being mostly on the periodic XXZ spin chain in odd size $N=2n+1$, the application will be some formulae expressing its ground state entries explicitly as coefficients of a multi-variable polynomial. This will allow us to settle some conjectures, including the calculation of the most antiferromagnetic component of the ground state.

The plan of the paper is as follows. In section 2, we introduce the various models involved.
In section 3, we formulate in the form of theorems the statements to be proved in what follows. In section 4, we discuss a certain relevant polynomial solution of the $q$KZ equation. Finally, in section 5, we take the homogeneous limit, prove the statements on the properties of the ground state components of XXZ spin chains at $\Delta=-1/2$ and Temperley-Lieb loop models formulated in Section~4, and comment on the implication for refined enumeration of Alternating Sign Matrices and Totally Symmetric Self-Complementary Plane Partitions.

\section{Six-vertex model and XXZ spin chain}

\subsection{Six-vertex model}

\begin{figure}[htb]
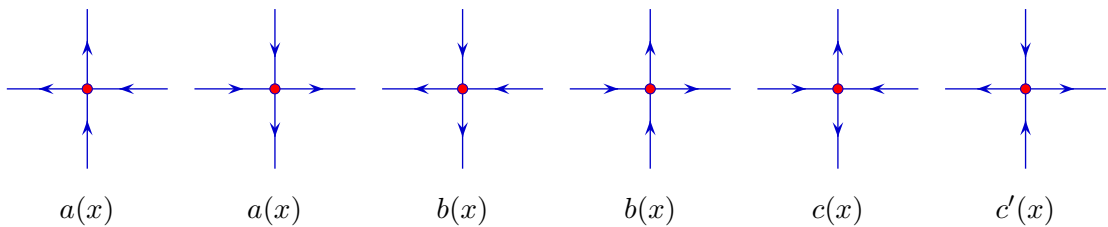

{\small \[
\begin{array}{cccccc}
\pspicture(0,0)(60,60)
  \psset{linecolor=MediumBlue}
  \arrowLine(30,30)(0,30)
  \arrowLine(30,30)(30,60)
  \arrowLine(60,30)(30,30)
  \arrowLine(30,0)(30,30)
  \pscircle*[linecolor=Red](30, 30){2}
  \pscircle[linewidth=.25bp](30, 30){2}
\endpspicture
&
\pspicture(0,0)(60,60)
  \psset{linecolor=MediumBlue}
  \arrowLine(0,30)(30,30)
  \arrowLine(30,60)(30,30)
  \arrowLine(30,30)(60,30)
  \arrowLine(30,30)(30,0)
  \pscircle*[linecolor=Red](30, 30){2}
  \pscircle[linewidth=.25bp](30, 30){2}
\endpspicture
&
\pspicture(0,0)(60,60)
  \psset{linecolor=MediumBlue}
  \arrowLine(30,30)(0,30)
  \arrowLine(30,60)(30,30)
  \arrowLine(60,30)(30,30)
  \arrowLine(30,30)(30,0)
  \pscircle*[linecolor=Red](30, 30){2}
  \pscircle[linewidth=.25bp](30, 30){2}
\endpspicture
&
\pspicture(0,0)(60,60)
  \psset{linecolor=MediumBlue}
  \arrowLine(0,30)(30,30)
  \arrowLine(30,30)(30,60)
  \arrowLine(30,30)(60,30)
  \arrowLine(30,0)(30,30)
  \pscircle*[linecolor=Red](30, 30){2}
  \pscircle[linewidth=.25bp](30, 30){2}
\endpspicture
&
\pspicture(0,0)(60,60)
  \psset{linecolor=MediumBlue}
  \arrowLine(0,30)(30,30)
  \arrowLine(30,30)(30,60)
  \arrowLine(60,30)(30,30)
  \arrowLine(30,30)(30,0)
  \pscircle*[linecolor=Red](30, 30){2}
  \pscircle[linewidth=.25bp](30, 30){2}
\endpspicture
&
\pspicture(0,0)(60,60)
  \psset{linecolor=MediumBlue}
  \arrowLine(30,30)(0,30)
  \arrowLine(30,60)(30,30)
  \arrowLine(30,30)(60,30)
  \arrowLine(30,0)(30,30)
  \pscircle*[linecolor=Red](30, 30){2}
  \pscircle[linewidth=.25bp](30, 30){2}
\endpspicture \\[.5em]
a(x) & a(x) & b(x) & b(x) & c(x) & c'(x)
\end{array}
\]}
\caption{The Boltzmann weights for the six-vertex model} \label{f:1}
\end{figure}

The six-vertex model is a statistical mechanics vertex model defined on a square lattice with $N$ vertical and $M$ horizontal rows. A state of the model is specified by a choice of the direction of each edge usually denoted by an arrow. The arrows obey the rule, called the ice condition, that at every vertex there are two arrows pointing in and two arrows pointing out. There are six possible configurations of arrows at each vertex, hence the name of the model. The Boltzmann weight of a vertex depends on its configuration and the value of the spectral parameter $x$ associated with the vertex as is given in Figure \ref{f:1},
where the functions $a(x)$, $b(x)$, $c(x)$ and $c'(x)$ are defined by the equalities\footnote{Note, that we use the so-called homogeneous gradation.}
\[
a(x) = \frac{q \, x - q^{-1}}{q - q^{-1} x}, \quad b(x) = \frac{x - 1}{q - q^{-1} x}, \quad c(x) = \frac{(q - q^{-1}) x}{q - q^{-1} x}, \quad c'(x) = \frac{q - q^{-1}}{q - q^{-1} x}.
\]
The parameter $q \ne \pm 1$ is a common parameter for all vertices. Often
the parameter
\[
\Delta = \frac{a^2(x)+b^2(x)-c(x)c'(x)}{2a(x) b(x)} =\frac{1}{2}(q+q^{-1})
\]
is used instead.
In the homogeneous case the same spectral parameter $x$ is associated with all vertices, while in the inhomogeneous case one associates variables $y_1$, $\ldots$, $y_M$ with the horizontal rows of the lattice and variables $z_1$, $\ldots$, $z_N$ with the vertical rows. A vertex at the intersection of horizontal row $p$ and vertical row $i$ acquires the spectral parameter equal to $y_p/z_i$.

Instead of orientation one can characterize the state of an edge by up or down arrows. Here we supply an edge with an up arrow if the edge orientation arrow points up or to the left and we supply it with a down arrow if the edge orientation arrow points down or to the right. It is convenient to assume that the states which do not satisfy the ice condition are also allowed but have the weight equal to zero and arrange the weights into a $4 \times 4$ matrix $R(x)$ whose rows are labeled by two indices, say $\alpha$ and $\mu$, taking the values $\uparrow$ and $\downarrow$, and whose columns are labeled by two indices, say $\beta$ and $\nu$, also taking the values $\uparrow$ and $\downarrow$. Assume that the correspondence of indices and edges is as follows
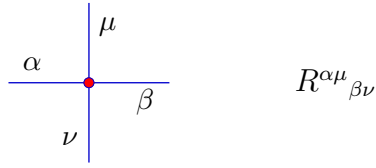
\[
\begin{pspicture}(0,0)(60,60)
  \small
  \psset{linecolor=MediumBlue}
  \psline(0,30)(60,30)
  \psline(30,0)(30,60)
  \pscircle*[linecolor=Red](30, 30){2}
  \pscircle[linewidth=.25bp](30, 30){2}
  \rput(9,37){$\alpha$}
  \rput(37,51){$\mu$}
  \rput(51,23){$\beta$}
  \rput(23,9){$\nu$}
\end{pspicture}  \qquad \qquad \raise 2.3em\hbox{$R^{\alpha \mu}{}_{\beta \nu}$}
\]
Choosing for pairs of indices the ordering $\uparrow \uparrow$, $\uparrow \downarrow$, $\downarrow \uparrow$, $\downarrow \downarrow$, we have
\[
R(x) = \left( \begin{array}{cccc}
a(x) &  &  &  \\
 & b(x) & c(x) &  \\
 & c'(x) & b(x) &  \\
 & & & a(x)
\end{array} \right),
\]
where only nonzero entries are presented.

We identify $R(x)$ with the linear operator $R_{1,2}(x)$ in the vector space $\bbC^2 \otimes \bbC^2$ defined via the equality
\[
R_{1,2}(x) e_{\nu_1} \otimes e_{\nu_2} = e_{\mu_1} \otimes e_{\mu_2} R^{\mu_1 \mu_2}{}_{\nu_1 \nu_2}(x),
\]
where $e_\mu$, $\mu = \uparrow, \downarrow$, are the elements of the standard basis of $\bbC^2$:
\[
e_\uparrow = \begin{pmatrix} 1 \\ 0 \end{pmatrix}, \qquad e_\downarrow = \begin{pmatrix} 0 \\ 1 \end{pmatrix}.
\]
In general, let $\mathcal A = \{\alpha_1, \ldots, \alpha_I\}$ be some ordered set of indices, and $V_\alpha$ for each $\alpha \in \mathcal A$ be a copy of the space $\bbC^2$. We denote by $R_{\alpha, \beta}(x)$ the linear operator in $V_{\alpha_1} \otimes \cdots \otimes V_{\alpha_I}$ acting as $R(x)$ in $V_\alpha$ and $V_\beta$ and identically in all other factors. In this situation we also denote by $P_{\alpha, \beta}$ the transposition
\[
P_{\alpha, \beta}(v_{\alpha_1} \otimes \ldots \otimes v_\alpha \otimes \ldots \otimes v_\beta \otimes \ldots \otimes v_{\alpha_I}) = v_{\alpha_1} \otimes \ldots \otimes v_\beta \otimes \ldots \otimes v_\alpha \otimes \ldots \otimes v_{\alpha_I}.
\]
It can be shown that in the space $V_1 \otimes V_2 \otimes V_3$ one has the Yang--Baxter equation
\begin{equation}
R_{1,2}(x_1/x_2) R_{1,3}(x_1/x_3) R_{2,3}(x_2/x_3) = R_{2,3}(x_2/x_3) R_{1,3}(x_1/x_3) R_{1,2}(x_1/x_2). \label{e:yb}
\end{equation}

The basic object of the statistical mechanics is the partition function of the system. One of the way to find it for a vertex model is to introduce the transfer matrices. The transfer matrix, associated with horizontal rows of the lattice is defined as
\[
T(y | z_1, \ldots, z_N) = \mathrm{tr}_0 [R_{0,1}(y/z_1) \ldots R_{0,N}(y/z_N)],
\]
where for each $i = 1, \ldots, N$ the operator $R_{0,i}(y/z_i)$ is an operator in $V_0 \otimes V_1 \otimes \ldots \otimes V_N $ and $\mathrm{tr}_0$ means the partial trace over $V_0$. Here we assume that toroidal boundary conditions are imposed. Using the Yang-Baxter equation (\ref{e:yb}) one can show that
\[
[T(y | z_1, \ldots, z_N), T(y' | z_1, \ldots, z_N)] = 0
\]
for any $y$ and $y'$. Therefore, for fixed $z_1$, $\ldots$, $z_N$ one can bring the transfer matrices $T(y | z_1, \ldots, z_N)$ to upper triangular form simultaneously for all values of $y$. Here the diagonal matrix elements of the resulting matrices are eigenvalues.

The partition function is related to the transfer matrix as follows
\[
Z(y_1, \ldots, y_M | z_1, \ldots, z_N) = \mathrm{tr} [ T(y_1 | z_1, \ldots, z_N) \ldots T(y_M | z_1, \ldots, z_N)].
\]
Bringing all the matrices under the trace to upper triangular form one reduces the calculation of the partition sum to the calculation of the eigenvalues of the transfer matrix. Note that in the thermodynamic limit only the largest eigenvalues of the transfer matrix contribute to the partition function.

There are two main methods to find eigenvalues and eigenvectors of the transfer matrix. The first one is the Bethe ansatz, which is not discussed here, and the second is the method of Baxter functional relations. In accordance with the latter method any eigenvalue $\lambda(y | z_1, \ldots, z_N)$ of the transfer matrix $T(y | z_1, \ldots, z_N)$ satisfies the relation
\begin{multline}
\lambda(y | z_1, \ldots, z_N) \rho(y | z_1, \ldots, z_N) \\
= \left[ \prod_{i=1}^N a(y/z_i) \right] \rho(q^{-2} y | z_1, \ldots, z_N) + \left[ \prod_{i=1}^N b(y/z_i) \right] \rho(q^2 y | z_1, \ldots, z_N), \label{e:fr}
\end{multline}
where $\rho(y | z_1, \ldots, z_N)$ is some function which is actually an eigenvalue of the so-called Baxter $Q$-operator $Q(y | z_1, \ldots, z_N)$.

In the present paper we discuss a special case where one can find an explicit solution of Eq.~(\ref{e:fr}). Namely, we consider the case where $q = \rme^{\pm 2 \pi \rmi/3}$. As was argued by Baxter\footnote{Actually Baxter considers the more general eight-vertex model, see in this respect the papers \cite{Str01a,BazMan05,BazMan06}.}~\cite{Bax89} in this case Eq.~(\ref{e:fr}) has a solution for
\begin{equation}
\lambda(y | z_1, \ldots, z_N) = \prod_{i=1}^N [a(y/z_i) + b(y/z_i)]. \label{e:se}
\end{equation}
As became clear afterwards, the corresponding eigenvector of the transfer matrix is nontrivial only if $N$ is odd. The explicit form of the function $\rho(y | z_1, \ldots, z_N)$ was found by Alcaraz and Stroganov \cite{AlcStr03} and earlier by Stroganov \cite{Str01} for the homogeneous case. Note that with the parameterization of the weights used in the present paper $a(x) + b(x) = 1$ if $q = \rme^{\pm 2 \pi \rmi/3}$. Hence, if the vector corresponding to the special eigenvalue of the transfer matrix given by Eq.~(\ref{e:se}) is nontrivial it is an eigenvector with the eigenvalue $1$. In the present papers we suggest and prove some integral representations for the components of the eigenvector $\Psi(z_1, \ldots, z_N)$ of the transfer matrix corresponding to the eigenvalue~$1$ for the case when $q = \rme^{\pm 2 \pi \rmi/3}$ and $N$ is odd.

\subsection{XXZ spin chain}

The six-vertex model is closely related to the XXZ model describing interaction of spin one-half particles arranged in a chain. Here each particle interacts with its nearest neighbors only. The Hamiltonian of the model has the form
\begin{equation}
H_{\mathrm{XXZ}}(\Delta) = - \frac{1}{2} \sum_{i=1}^{N} \left[ \sigma_i^{x} \sigma_{i+1}^{x} + \sigma_i^{y} \sigma_{i+1}^{y} + \Delta \, \sigma_i^z \sigma_{i+1}^z \right],
\label{H1}
\end{equation}
where $\Delta$ is the so-called anisotropy parameter, and $\sigma^x_i$, $\sigma^y_i$ $\sigma^z_i$ are the operators describing spin degrees of freedom of the particle at site $i$ of the chain. We assume that periodic boundary conditions are imposed.

In the homogeneous limit where all $z_i$ are equal to $1$ for general $q$ the logarithmic derivative of the transfer matrix of the six-vertex model $T(x) = T(x | 1, \ldots, 1)$ at $x = 1$ is related to the Hamiltonian $H_{\mathrm{XXZ}}(\Delta)$ as
\begin{equation}
\left[ T^{-1}(x) \frac{\rmd T(x)}{\rmd x} \right]_{x=1} = - \frac{1}{q - q^{-1}} \left[ H_{\mathrm{XXZ}}(\Delta) - \frac{3 N}{2} \Delta \right], \label{e:ld}
\end{equation}
where $\Delta = (q + q^{-1})/2$. In the case when $q = \rme^{\pm 2\pi\rmi/3}$ one has $q + q^{-1} = -1$ and $\Delta$ is equal to $-1/2$. Therefore, in this case the homogeneous limit $\psi$ of the eigenvector $\Psi(z_1, \ldots, z_N)$ of the transfer matrix with the eigenvalue $1$ is an eigenvector of the Hamiltonian $H_{\mathrm{XXZ}}(-1/2)$ with the eigenvalue $-3N/4$.

It is well-known that one can look for the eigenvectors of the transfer matrix of the six-vertex model and the Hamiltonian of the XXZ spin chain in the sectors spanned by the basis vectors with fixed number of down or up arrows. For the case of $K$ down arrows we denote the corresponding basis vectors by $e_{a_1, \ldots, a_K}$, where $a_1$, $\ldots$, $a_K$ are positions of down arrows. It is natural to assume that $1 \le a_1 < \ldots < a_K \le N$. The discussed eigenvectors of the transfer matrix and $H_{\mathrm{XXZ}}(-1/2)$ belongs to the sector with $n$ down arrows if $N = 2n+1$. Actually, it has a companion with the same eigenvalue in the sector with $n+1$ down arrows which can be obtained by the transformation reversing direction of arrows. This transformation commutes with the transfer matrix and $H_{\mathrm{XXZ}}(\Delta)$.

The eigenvector of the Hamiltonian $H_{\mathrm{XXZ}}(-1/2)$ with the eigenvalue $-3N/4$ for an odd number of sites was investigated numerically by Razumov and Stroganov~\cite{RazStr01}. They formulated a few conjectures about the properties of the components of this vector. Some of this conjectures have been proved already \cite{KitMaiSlaTer02a, KitMaiSlaTer02b, DiFZinZub06, Pas06}, and some of them have been generalized to the case of different boundary conditions \cite{BatdeGNie01, RazStr01b}. In particular, it was shown that the considered vector is the ground state of the Hamiltonian $H_{\mathrm{XXZ}}(-1/2)$ \cite{YanFen04, VenWos06}.

\subsection{Temperley--Lieb loop model} \label{s:2.3}

\newcommand{\bra}[1]{\langle #1|}
\newcommand{\ket}[1]{|#1\rangle}
\newcommand{\braket}[2]{\langle #1|#2\rangle}
Although this is not the main focus of this work, we shall briefly describe here a related model of loops. This will allow us to derive interesting new connections between the discussed models, and also to compare our results with those of the paper \cite{DiFZin07}.

\begin{figure}[htb]
\[
\begin{pspicture}(-40, -40)(40, 40)
\SpecialCoor
\pscircle(0,0){30}
\rput(37; 210){$\scriptscriptstyle 1$}
\rput(37; 270){$\scriptscriptstyle 2$}
\rput(37; 330){$\scriptscriptstyle 3$}
\rput(37; 30){$\scriptscriptstyle 4$}
\rput(37; 90){$\scriptscriptstyle 5$}
\rput(37; 150){$\scriptscriptstyle 6$}
\psbezier[linecolor=MediumBlue,linewidth=1bp](30; 210)(18; 220)(18; 260)(30; 270)
\psline[linecolor=MediumBlue,linewidth=1bp](30; 330)(30; 150)
\psbezier[linecolor=MediumBlue,linewidth=1bp](30; 30)(18; 40)(18; 80)(30; 90)
\pscircle*[linecolor=Red](30; 210){2}
\pscircle*[linecolor=Red](30; 270){2}
\pscircle*[linecolor=Red](30; 330){2}
\pscircle*[linecolor=Red](30; 30){2}
\pscircle*[linecolor=Red](30; 90){2}
\pscircle*[linecolor=Red](30; 150){2}
\pscircle[linewidth=.25bp](30; 210){2}
\pscircle[linewidth=.25bp](30; 270){2}
\pscircle[linewidth=.25bp](30; 330){2}
\pscircle[linewidth=.25bp](30; 30){2}
\pscircle[linewidth=.25bp](30; 90){2}
\pscircle[linewidth=.25bp](30; 150){2}
\end{pspicture}
\begin{pspicture}(-40, -40)(40, 40)
\SpecialCoor
\pscircle(0,0){30}
\rput(37; 210){$\scriptscriptstyle 1$}
\rput(37; 270){$\scriptscriptstyle 2$}
\rput(37; 330){$\scriptscriptstyle 3$}
\rput(37; 30){$\scriptscriptstyle 4$}
\rput(37; 90){$\scriptscriptstyle 5$}
\rput(37; 150){$\scriptscriptstyle 6$}
\psbezier[linecolor=MediumBlue,linewidth=1bp](30; 270)(18; 280)(18; 320)(30; 330)
\psline[linecolor=MediumBlue,linewidth=1bp](30; 30)(30; 210)
\psbezier[linecolor=MediumBlue,linewidth=1bp](30; 90)(18; 100)(18; 140)(30; 150)
\pscircle*[linecolor=Red](30; 210){2}
\pscircle*[linecolor=Red](30; 270){2}
\pscircle*[linecolor=Red](30; 330){2}
\pscircle*[linecolor=Red](30; 30){2}
\pscircle*[linecolor=Red](30; 90){2}
\pscircle*[linecolor=Red](30; 150){2}
\pscircle[linewidth=.25bp](30; 210){2}
\pscircle[linewidth=.25bp](30; 270){2}
\pscircle[linewidth=.25bp](30; 330){2}
\pscircle[linewidth=.25bp](30; 30){2}
\pscircle[linewidth=.25bp](30; 90){2}
\pscircle[linewidth=.25bp](30; 150){2}
\end{pspicture}
\begin{pspicture}(-40, -40)(40, 40)
\SpecialCoor
\pscircle(0,0){30}
\rput(37; 210){$\scriptscriptstyle 1$}
\rput(37; 270){$\scriptscriptstyle 2$}
\rput(37; 330){$\scriptscriptstyle 3$}
\rput(37; 30){$\scriptscriptstyle 4$}
\rput(37; 90){$\scriptscriptstyle 5$}
\rput(37; 150){$\scriptscriptstyle 6$}
\psline[linecolor=MediumBlue,linewidth=1bp](30; 270)(30; 90)
\psbezier[linecolor=MediumBlue,linewidth=1bp](30; 330)(18; 340)(18; 20)(30; 30)
\psbezier[linecolor=MediumBlue,linewidth=1bp](30; 150)(18; 160)(18; 200)(30; 210)
\pscircle*[linecolor=Red](30; 210){2}
\pscircle*[linecolor=Red](30; 270){2}
\pscircle*[linecolor=Red](30; 330){2}
\pscircle*[linecolor=Red](30; 30){2}
\pscircle*[linecolor=Red](30; 90){2}
\pscircle*[linecolor=Red](30; 150){2}
\pscircle[linewidth=.25bp](30; 210){2}
\pscircle[linewidth=.25bp](30; 270){2}
\pscircle[linewidth=.25bp](30; 330){2}
\pscircle[linewidth=.25bp](30; 30){2}
\pscircle[linewidth=.25bp](30; 90){2}
\pscircle[linewidth=.25bp](30; 150){2}
\end{pspicture}
\begin{pspicture}(-40, -40)(40, 40)
\SpecialCoor
\pscircle(0,0){30}
\rput(37; 210){$\scriptscriptstyle 1$}
\rput(37; 270){$\scriptscriptstyle 2$}
\rput(37; 330){$\scriptscriptstyle 3$}
\rput(37; 30){$\scriptscriptstyle 4$}
\rput(37; 90){$\scriptscriptstyle 5$}
\rput(37; 150){$\scriptscriptstyle 6$}
\psbezier[linecolor=MediumBlue,linewidth=1bp](30; 210)(18; 220)(18; 260)(30; 270)
\psbezier[linecolor=MediumBlue,linewidth=1bp](30; 330)(18; 340)(18; 20)(30; 30)
\psbezier[linecolor=MediumBlue,linewidth=1bp](30; 90)(18; 100)(18; 140)(30; 150)
\pscircle*[linecolor=Red](30; 210){2}
\pscircle*[linecolor=Red](30; 270){2}
\pscircle*[linecolor=Red](30; 330){2}
\pscircle*[linecolor=Red](30; 30){2}
\pscircle*[linecolor=Red](30; 90){2}
\pscircle*[linecolor=Red](30; 150){2}
\pscircle[linewidth=.25bp](30; 210){2}
\pscircle[linewidth=.25bp](30; 270){2}
\pscircle[linewidth=.25bp](30; 330){2}
\pscircle[linewidth=.25bp](30; 30){2}
\pscircle[linewidth=.25bp](30; 90){2}
\pscircle[linewidth=.25bp](30; 150){2}
\end{pspicture}
\begin{pspicture}(-40, -40)(40, 40)
\SpecialCoor
\pscircle(0,0){30}
\rput(37; 210){$\scriptscriptstyle 1$}
\rput(37; 270){$\scriptscriptstyle 2$}
\rput(37; 330){$\scriptscriptstyle 3$}
\rput(37; 30){$\scriptscriptstyle 4$}
\rput(37; 90){$\scriptscriptstyle 5$}
\rput(37; 150){$\scriptscriptstyle 6$}
\psbezier[linecolor=MediumBlue,linewidth=1bp](30; 270)(18; 280)(18; 320)(30; 330)
\psbezier[linecolor=MediumBlue,linewidth=1bp](30; 30)(18; 40)(18; 80)(30; 90)
\psbezier[linecolor=MediumBlue,linewidth=1bp](30; 150)(18; 160)(18; 200)(30; 210)
\pscircle*[linecolor=Red](30; 210){2}
\pscircle*[linecolor=Red](30; 270){2}
\pscircle*[linecolor=Red](30; 330){2}
\pscircle*[linecolor=Red](30; 30){2}
\pscircle*[linecolor=Red](30; 90){2}
\pscircle*[linecolor=Red](30; 150){2}
\pscircle[linewidth=.25bp](30; 210){2}
\pscircle[linewidth=.25bp](30; 270){2}
\pscircle[linewidth=.25bp](30; 330){2}
\pscircle[linewidth=.25bp](30; 30){2}
\pscircle[linewidth=.25bp](30; 90){2}
\pscircle[linewidth=.25bp](30; 150){2}
\end{pspicture}
\]
\caption{The link patterns forming the canonical basis of the Temperley--Lieb loop model for $2n = 6$.} \label{f:2}
\end{figure}
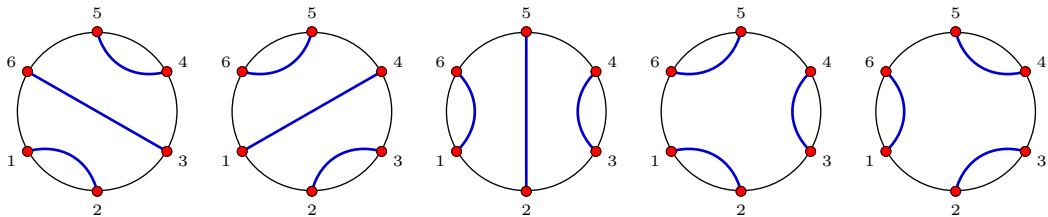

We assume that the size of the system is even, equal to $2n$. The state space of the model is the free vector space generated by the set $\Pi_{2n}$ formed by {\em link patterns} $\pi$, that is nonintersecting planar pairings of $2n$ points regularly distributed on a circle. The dimension of this space is equal to the Catalan number $\frac{1}{n+1} \binom{2n}{n}$. For example, in size $2n=6$, there are 5 link patterns given in Figure \ref{f:2}.
It is sometimes convenient to identify a link pattern $\pi$ with the involution that exchanges paired points, so that such pairs are of the form $(i,\pi(i))$.

Define linear operators $e_i$, $i=1, \ldots, 2n$, which act on link
patterns according to the following rule: either the points $i$ and
$i+1$ (mod $2n$) are already connected in the initial link pattern, in
which case the link pattern is simply multiplied by $\tau = -q - q^{-1}$; or they are not, in which case the lines arriving at points $i$, $i+1$ are reconnected and a new arch pairs $i$ and $i+1$ in the new link pattern. For example we have
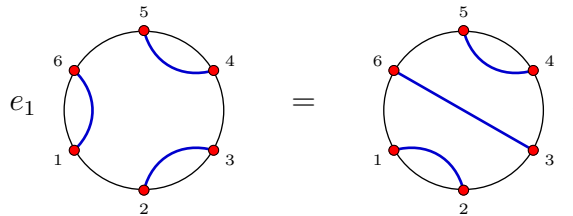
\[
e_1 \lower 40bp \hbox{\begin{pspicture}(-40, -40)(40, 40)
\SpecialCoor
\pscircle(0,0){30}
\rput(37; 210){$\scriptscriptstyle 1$}
\rput(37; 270){$\scriptscriptstyle 2$}
\rput(37; 330){$\scriptscriptstyle 3$}
\rput(37; 30){$\scriptscriptstyle 4$}
\rput(37; 90){$\scriptscriptstyle 5$}
\rput(37; 150){$\scriptscriptstyle 6$}
\psbezier[linecolor=MediumBlue,linewidth=1bp](30; 270)(18; 280)(18; 320)(30; 330)
\psbezier[linecolor=MediumBlue,linewidth=1bp](30; 30)(18; 40)(18; 80)(30; 90)
\psbezier[linecolor=MediumBlue,linewidth=1bp](30; 150)(18; 160)(18; 200)(30; 210)
\pscircle*[linecolor=Red](30; 210){2}
\pscircle*[linecolor=Red](30; 270){2}
\pscircle*[linecolor=Red](30; 330){2}
\pscircle*[linecolor=Red](30; 30){2}
\pscircle*[linecolor=Red](30; 90){2}
\pscircle*[linecolor=Red](30; 150){2}
\pscircle[linewidth=.25bp](30; 210){2}
\pscircle[linewidth=.25bp](30; 270){2}
\pscircle[linewidth=.25bp](30; 330){2}
\pscircle[linewidth=.25bp](30; 30){2}
\pscircle[linewidth=.25bp](30; 90){2}
\pscircle[linewidth=.25bp](30; 150){2}
\end{pspicture}} \quad = \quad
\lower 40bp \hbox{\begin{pspicture}(-40, -40)(40, 40)
\SpecialCoor
\pscircle(0,0){30}
\rput(37; 210){$\scriptscriptstyle 1$}
\rput(37; 270){$\scriptscriptstyle 2$}
\rput(37; 330){$\scriptscriptstyle 3$}
\rput(37; 30){$\scriptscriptstyle 4$}
\rput(37; 90){$\scriptscriptstyle 5$}
\rput(37; 150){$\scriptscriptstyle 6$}
\psbezier[linecolor=MediumBlue,linewidth=1bp](30; 210)(18; 220)(18; 260)(30; 270)
\psline[linecolor=MediumBlue,linewidth=1bp](30; 330)(30; 150)
\psbezier[linecolor=MediumBlue,linewidth=1bp](30; 30)(18; 40)(18; 80)(30; 90)
\pscircle*[linecolor=Red](30; 210){2}
\pscircle*[linecolor=Red](30; 270){2}
\pscircle*[linecolor=Red](30; 330){2}
\pscircle*[linecolor=Red](30; 30){2}
\pscircle*[linecolor=Red](30; 90){2}
\pscircle*[linecolor=Red](30; 150){2}
\pscircle[linewidth=.25bp](30; 210){2}
\pscircle[linewidth=.25bp](30; 270){2}
\pscircle[linewidth=.25bp](30; 330){2}
\pscircle[linewidth=.25bp](30; 30){2}
\pscircle[linewidth=.25bp](30; 90){2}
\pscircle[linewidth=.25bp](30; 150){2}
\end{pspicture}}.
\]
One can easily show that the operators $e_i$ satisfy defining relations of the Temperley--Lieb algebra \cite{TemLie}.

Finally the Hamiltonian is defined by $H_{\mathrm{TL}}=\sum_{i=1}^{2n} e_i$. When $q = \rme^{\pm 2 \pi \rmi/3}$ one has $\tau = 1$. It is easy to check that in this case the dual of $H_{\mathrm{TL}}$ has an eigenvector with all components relative to the dual of the basis $\Pi_{2n}$ equal to $1$ and with eigenvalue $2n$. Using the Perron--Frobenius theorem, one can show that this is the largest eigenvalue. With an opposite sign convention for $H_{\mathrm{TL}}$, this would be the ground state eigenvalue. Normalize the corresponding eigenvector
\[
\xi=\sum_{\pi \in \Pi_{2n}} \pi \xi_\pi
\]
of $H_{\mathrm{TL}}$ so that the smallest components, corresponding to the link patterns of the type of the first three patterns in Figure \ref{f:2}, is set to $1$. In this case the components $\xi_\pi$ possess remarkable combinatorial properties, see in particular the paper \cite{RazStr04}.

An additional remark is in order: it is known that this model of loops can be mapped into
a {\em twisted}\/ XXZ spin chain of the same even size. This is not the odd-sized XXZ model that
is considered in the present work. Nevertheless we shall find below some non-trivial connections
between odd-sized XXZ spin chain and even-sized loop model.

\section{Formulation of the main results}

We now formulate in the form of theorems three properties of the ground state components of the models above that will be proved in Section \ref{s:5}. The first two of them were conjectured previously in the paper \cite{RazStr01}, and the third one is new. Denote by $\psi$ the ground state eigenvector of the XXZ spin chain of size $2n+1$ at $\Delta=-1/2$, and by $\xi$ the ground state eigenvector of the Temperley--Lieb loop model of size $2n$ at $\tau=1$. Recall that $\psi_{a_1,\ldots,a_n}$ is the component of $\psi$ with down arrows at locations $a_1,\ldots,a_n$, and $\xi_\pi$ is the component of $\xi$ corresponding to the link pattern $\pi$.

\subsection{Integerness}

The first result reads:
\begin{theorem} \label{c:1}
If we choose the normalization of the vector $\psi$ so that $\psi_{1,2,\ldots,n}=1$
then all other components $\psi_{a_1,\ldots,a_n}$ are integer.
\end{theorem}

This is one of the conjectures of the paper \cite{RazStr01}. In fact, it is not hard to check that the Perron--Frobenius theorem can be applied here, see, for example, \cite[Theorem 4]{YanYan}, so that with this normalization the components are also positive.

A similar conjecture exists for the Temperley--Lieb loop model, but will not be
addressed here.

\subsection{Some partial sums and refined ASM enumeration} \label{s:3.2}

An alternating sign matrix (ASM) is a square matrix with entries in $\{
-1,0,+1\}$ such that in every row and column, the sequence of $\pm 1$
is alternatingly $+1$ and $-1$ starting and ending with $+1$.
Denote by $A(n)$ the number of ASMs of size $n$, and by $A(n,r)$ the
number of ASMs of size $n$ whose (unique) $+1$ in the first row is at
column $r$. These numbers have been at the center of a great deal of activity, see the book \cite{Bres}. As was conjectured by Mills, Robbins, and Rumsey \cite{MilRobRum82,MilRobRum83} and proved by Zeilberger \cite{Zei96a,Zei96b} they are given by the formulae
\begin{gather*}
A(n) = \frac{1! \, 4! \, 7! \cdots (3n-2)!}{n! \, (n+1)! \cdots (2n-1)!}, \\
A(n,r) = A(n) \frac{\binom{n+r-2}{n-1} \binom{2n-1-r}{n-1}}{\binom{3n-2}{n-1}}.
\end{gather*}

Let us now return to the ground state eigenvector $\psi$. In the paper \cite{RazStr06},
Razumov and Stroganov studied certain components $\psi_{a_1,\ldots,a_n}$ such that $a_1=1$ or $2$, \dots, $a_\ell=2\ell-1$ or $2\ell$, \dots, $a_n=2n-1$ or $2n$. They made the following observation, based on numerical evidence:
\begin{equation}
\sum_{\varepsilon_1, \ldots, \varepsilon_n \in \{0,1\}} \alpha^{\sum_{\ell=1}^n
  \varepsilon_\ell} \, \psi_{1 + \varepsilon_1, 3 + \varepsilon_2, \ldots,
  2n-1 + \varepsilon_n} = \sum_{r=1}^{n+1} \alpha^{r-1} A(n+1,r).
\label{e:refasmx}
\end{equation}
Using the coordinate Bethe Ansatz techniques, they managed to prove the following identity:
\begin{equation}
\frac{1}{\psi_{1, 3, \ldots, 2n-1}} \sum_{\varepsilon_1, \ldots, \varepsilon_n \in \{0,1\}} \alpha^{\sum_{\ell=1}^n \varepsilon_\ell} \, \psi_{1 + \varepsilon_1, 3 + \varepsilon_2, \ldots, 2n-1 + \varepsilon_n} = \frac{1}{A(n)} \sum_{r=1}^{n+1} \alpha^{r-1} A(n + 1,r),
\label{e:prerefasmx}
\end{equation}
We shall use this identity below. For now, what is important is that the proof of the original observation (\ref{e:refasmx}) is reduced to the proof of the following result:
\begin{theorem} \label{c:2}
If we choose the normalization of the vector $\psi$ so that $\psi_{1,2,\ldots,n}=1$
then
$\psi_{1,3, \ldots, 2n-1} = A(n)$.
\end{theorem}
Note that, in view of the elementary relation $A(n+1,1)=A(n)$, this is nothing but the special case $\alpha=0$ of Eq.~(\ref{e:refasmx}). In fact this was conjectured much earlier, in the paper~\cite{RazStr01}, where it was also observed that $\psi_{1,3,\ldots,2n-1}$ is the largest component.

\subsection{From the XXZ spin chain ground state to the Temperley--Lieb loop model ground state} \label{s:3.3}

Let us now consider the components obtained from the component $\psi_{1,3,\ldots,2n-1}$ by increasing one of the indices by $1$. This is best explained by an example. In size $N = 2n+1 = 9$, they are the components
\[
\psi_{2, 3, 5, 7} = 17,
\qquad
\psi_{1, 4, 5, 7} = 21,
\qquad
\psi_{1, 3, 6, 7} = 25,
\qquad
\psi_{1,3,5,8}=42.
\]
The last component is nothing but the component $\psi_{1,3,5,7} = A(4) = 42$, which is obvious because they are obtained from each other by rotation.

Now, amazingly the same quantities appear in the Temperley-Lieb loop model of size $2n$, that is equal to $8$ in our example.
Explicitly, group the link patterns $\pi$ according to the point $\pi(1)$ to which $1$ is paired.
Consider the partial sums of components $\xi_{(1,a)}=\sum_{\pi:\, \pi(1)=a} \xi_\pi$.
For $2n = 8$ we find
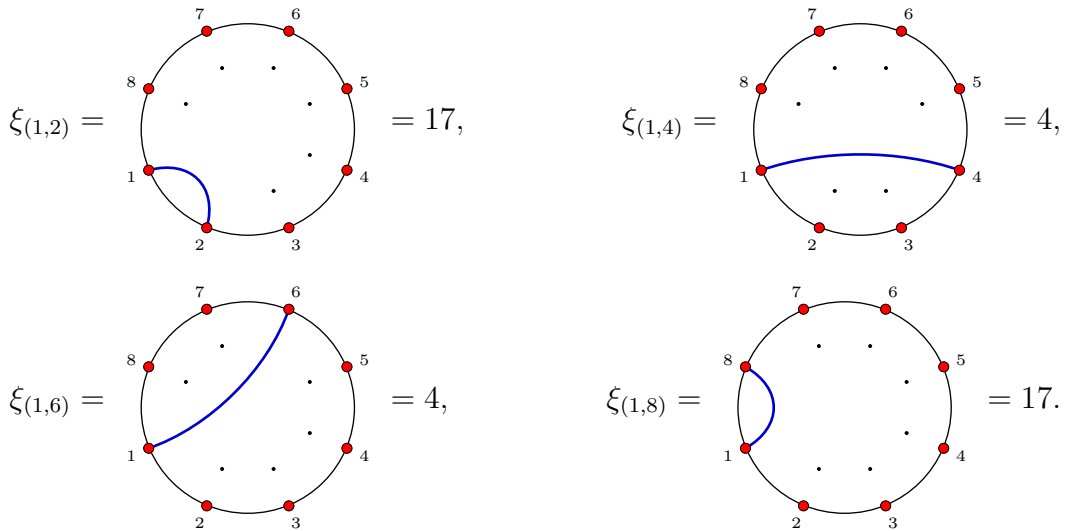
\begin{align*}
& \xi_{(1,2)} = \lower 50bp \hbox{\begin{pspicture}(-50, -50)(50, 50)
\SpecialCoor
\pscircle(0,0){40}
\rput(47; 202.5){$\scriptscriptstyle 1$}
\rput(47; 247.5){$\scriptscriptstyle 2$}
\rput(47; 292.5){$\scriptscriptstyle 3$}
\rput(47; 337.5){$\scriptscriptstyle 4$}
\rput(47; 22.5){$\scriptscriptstyle 5$}
\rput(47; 67.5){$\scriptscriptstyle 6$}
\rput(47; 112.5){$\scriptscriptstyle 7$}
\rput(47; 157.5){$\scriptscriptstyle 8$}
\psbezier[linecolor=MediumBlue,linewidth=1bp](40; 202.5)(24; 207.5)(24; 242.5)(40; 247.5)
\pscircle*[linecolor=Red](40; 202.5){2}
\pscircle*[linecolor=Red](40; 247.5){2}
\pscircle*[linecolor=Red](40; 292.5){2}
\pscircle*[linecolor=Red](40; 337.5){2}
\pscircle*[linecolor=Red](40; 22.5){2}
\pscircle*[linecolor=Red](40; 67.5){2}
\pscircle*[linecolor=Red](40; 112.5){2}
\pscircle*[linecolor=Red](40; 157.5){2}
\pscircle[linewidth=.25bp](40; 202.5){2}
\pscircle[linewidth=.25bp](40; 247.5){2}
\pscircle[linewidth=.25bp](40; 292.5){2}
\pscircle[linewidth=.25bp](40; 337.5){2}
\pscircle[linewidth=.25bp](40; 22.5){2}
\pscircle[linewidth=.25bp](40; 67.5){2}
\pscircle[linewidth=.25bp](40; 112.5){2}
\pscircle[linewidth=.25bp](40; 157.5){2}
\psdots[dotsize=1.5bp] (25; 157.5) (25; 112.5) (25; 67.5) (25; 22.5) (25; 337.5) (25; 292.5)
\end{pspicture}} = 17, \qquad &
\xi_{(1,4)} = \lower 50bp \hbox{\begin{pspicture}(-50, -50)(50, 50)
\SpecialCoor
\pscircle(0,0){40}
\rput(47; 202.5){$\scriptscriptstyle 1$}
\rput(47; 247.5){$\scriptscriptstyle 2$}
\rput(47; 292.5){$\scriptscriptstyle 3$}
\rput(47; 337.5){$\scriptscriptstyle 4$}
\rput(47; 22.5){$\scriptscriptstyle 5$}
\rput(47; 67.5){$\scriptscriptstyle 6$}
\rput(47; 112.5){$\scriptscriptstyle 7$}
\rput(47; 157.5){$\scriptscriptstyle 8$}
\psbezier[linecolor=MediumBlue,linewidth=1bp](40; 202.5)(16; 207.5)(16; 332.5)(40; 337.5)
\pscircle*[linecolor=Red](40; 202.5){2}
\pscircle*[linecolor=Red](40; 247.5){2}
\pscircle*[linecolor=Red](40; 292.5){2}
\pscircle*[linecolor=Red](40; 337.5){2}
\pscircle*[linecolor=Red](40; 22.5){2}
\pscircle*[linecolor=Red](40; 67.5){2}
\pscircle*[linecolor=Red](40; 112.5){2}
\pscircle*[linecolor=Red](40; 157.5){2}
\pscircle[linewidth=.25bp](40; 202.5){2}
\pscircle[linewidth=.25bp](40; 247.5){2}
\pscircle[linewidth=.25bp](40; 292.5){2}
\pscircle[linewidth=.25bp](40; 337.5){2}
\pscircle[linewidth=.25bp](40; 22.5){2}
\pscircle[linewidth=.25bp](40; 67.5){2}
\pscircle[linewidth=.25bp](40; 112.5){2}
\pscircle[linewidth=.25bp](40; 157.5){2}
\psdots[dotsize=1.5bp] (25; 157.5) (25; 112.5) (25; 67.5) (25; 22.5) (25; 292.5) (25; 247.5)
\end{pspicture}} = 4,
\\
& \xi_{(1,6)} = \lower 50bp \hbox{\begin{pspicture}(-50, -50)(50, 50)
\SpecialCoor
\pscircle(0,0){40}
\rput(47; 202.5){$\scriptscriptstyle 1$}
\rput(47; 247.5){$\scriptscriptstyle 2$}
\rput(47; 292.5){$\scriptscriptstyle 3$}
\rput(47; 337.5){$\scriptscriptstyle 4$}
\rput(47; 22.5){$\scriptscriptstyle 5$}
\rput(47; 67.5){$\scriptscriptstyle 6$}
\rput(47; 112.5){$\scriptscriptstyle 7$}
\rput(47; 157.5){$\scriptscriptstyle 8$}
\psbezier[linecolor=MediumBlue,linewidth=1bp](40; 202.5)(16; 207.5)(16; 62.5)(40; 67.5)
\pscircle*[linecolor=Red](40; 202.5){2}
\pscircle*[linecolor=Red](40; 247.5){2}
\pscircle*[linecolor=Red](40; 292.5){2}
\pscircle*[linecolor=Red](40; 337.5){2}
\pscircle*[linecolor=Red](40; 22.5){2}
\pscircle*[linecolor=Red](40; 67.5){2}
\pscircle*[linecolor=Red](40; 112.5){2}
\pscircle*[linecolor=Red](40; 157.5){2}
\pscircle[linewidth=.25bp](40; 202.5){2}
\pscircle[linewidth=.25bp](40; 247.5){2}
\pscircle[linewidth=.25bp](40; 292.5){2}
\pscircle[linewidth=.25bp](40; 337.5){2}
\pscircle[linewidth=.25bp](40; 22.5){2}
\pscircle[linewidth=.25bp](40; 67.5){2}
\pscircle[linewidth=.25bp](40; 112.5){2}
\pscircle[linewidth=.25bp](40; 157.5){2}
\psdots[dotsize=1.5bp] (25; 157.5) (25; 112.5) (25; 22.5) (25; 337.5) (25; 292.5) (25; 247.5)
\end{pspicture}} = 4, &
\xi_{(1,8)} = \lower 50bp \hbox{\begin{pspicture}(-50, -50)(50, 50)
\SpecialCoor
\pscircle(0,0){40}
\rput(47; 202.5){$\scriptscriptstyle 1$}
\rput(47; 247.5){$\scriptscriptstyle 2$}
\rput(47; 292.5){$\scriptscriptstyle 3$}
\rput(47; 337.5){$\scriptscriptstyle 4$}
\rput(47; 22.5){$\scriptscriptstyle 5$}
\rput(47; 67.5){$\scriptscriptstyle 6$}
\rput(47; 112.5){$\scriptscriptstyle 7$}
\rput(47; 157.5){$\scriptscriptstyle 8$}
\psbezier[linecolor=MediumBlue,linewidth=1bp](40; 202.5)(24; 197.5)(24; 162.5)(40; 157.5)
\pscircle*[linecolor=Red](40; 202.5){2}
\pscircle*[linecolor=Red](40; 247.5){2}
\pscircle*[linecolor=Red](40; 292.5){2}
\pscircle*[linecolor=Red](40; 337.5){2}
\pscircle*[linecolor=Red](40; 22.5){2}
\pscircle*[linecolor=Red](40; 67.5){2}
\pscircle*[linecolor=Red](40; 112.5){2}
\pscircle*[linecolor=Red](40; 157.5){2}
\pscircle[linewidth=.25bp](40; 202.5){2}
\pscircle[linewidth=.25bp](40; 247.5){2}
\pscircle[linewidth=.25bp](40; 292.5){2}
\pscircle[linewidth=.25bp](40; 337.5){2}
\pscircle[linewidth=.25bp](40; 22.5){2}
\pscircle[linewidth=.25bp](40; 67.5){2}
\pscircle[linewidth=.25bp](40; 112.5){2}
\pscircle[linewidth=.25bp](40; 157.5){2}
\psdots[dotsize=1.5bp] (25; 112.5) (25; 67.5) (25; 22.5) (25; 337.5) (25; 292.5) (25; 247.5)
\end{pspicture}} = 17.
\end{align*}
Calculating successive partial sums of the first $1$, $2$, $3$ and $4$ of the obtained numbers, we reproduce exactly the components of $\psi$ above. We are thus led to the following statement, which is proved in what follows:
\begin{theorem} \label{c:3}
There exist linear relations between the normalized ground state of the XXZ spin chain in size $2n+1$ and the normalized ground state of the Temperley--Lieb loop model in size $2n$ of the form
\[
\psi_{1,3,\ldots,2k-3,2k,2k+1,\ldots,2n-1}=\sum_{m=1}^k \xi_{(1,2m)}, \qquad k = 1, \ldots, n.
\]
\end{theorem}
Note that at $k=n$, the above identity states that the component $\psi_{1,3,\ldots,2n-3,2n}$ is equal to the sum of all components of the loop model. We have already mentioned that the former is conjectured to be $A(n)$. That the latter is also equal to $A(n)$ was conjectured by Batchelor, de~Gier and Nienhuis \cite{BatdeGNie01} and proved by Di~Francesco and Zinn-Justin \cite{DiFZin05a}.

\section{Polynomial solutions of $q$KZ equation}

\subsection{Integral formulae}

Being inspired by the integral formulae given, for example, in the book by Jimbo and Miwa \cite{JimMiw95} let us consider the following integral
\begin{multline}
\Psi_{a_1, \ldots, a_n}(z_1,\ldots, z_N) \\
= \prod_{1 \le i < j \le N}(q\,z_i-q^{-1}z_j) \oint \cdots \oint \prod_{\ell=1}^{n} \frac{\rmd w_\ell}{2 \pi \rmi} \Phi_{a_1, \ldots, a_n}(w_1, \ldots, w_n | z_1, \ldots, z_N), \label{e:b}
\end{multline}
where $a_1, \ldots, a_n$ is a sequence of positive integers such that $1 \le a_1 < a_2 < \ldots < a_n \le N=2n+1$, and the function $\Phi_{a_1, \ldots, a_n}(w_1, \ldots, w_n | z_1, \ldots, z_N)$ has the form
\begin{multline*}
\Phi_{a_1, \ldots, a_n}(w_1, \ldots, w_n | z_1, \ldots, z_N)
\\
= (q - q^{-1})^n \prod_{\ell=1}^n z_{a_\ell} \frac{\prod_{\ell=1}^{n} w_\ell \prod_{1 \le \ell < m \le n} [(w_m - w_\ell) (q\, w_\ell - q^{-1} w_m)]}{\prod_{\ell=1}^n \left[ \prod_{1 \le i \le a_\ell}(w_\ell - z_i) \prod_{a_\ell \le i \le N}(q\, w_\ell - q^{-1} z_i) \right]}.
\end{multline*}
The variables $w_1$, $\ldots$, $w_n$ and $z_1$, $\ldots$, $z_N$ are complex, and the integration contours surround the poles at $w_\ell = z_i$, but not at $w_\ell = q^{-2}z_i$.

It is also necessary for our purposes
to consider an alternative integral formula, which involves $n+1$ integrations:
\begin{multline}
\overline{\Psi}_{b_1, \ldots, b_{n+1}}(z_1,\ldots, z_N) \\
= \prod_{1 \le i < j \le N}(q \, z_i - q^{-1}z_j) \oint \cdots \oint \prod_{\ell=1}^{n+1} \frac{\rmd w_\ell}{2 \pi \rmi} \overline{\Phi}_{b_1, \ldots, b_{n+1}}(w_1, \ldots, w_{n+1} | z_1, \ldots, z_N), \label{e:a}
\end{multline}
where $1 \le b_1 < b_2 \ldots < b_{n+1} \le N$, the function $\overline{\Phi}_{b_1, \ldots, b_{n+1}}(w_1, \ldots, w_{n+1} | z_1, \ldots, z_N)$ has the form
\begin{multline*}
\overline{\Phi}_{b_1, \ldots, b_{n+1}}(w_1, \ldots, w_{n+1} | z_1, \ldots, z_N)
\\
= (q - q^{-1})^n \prod_{i=1}^N z_i \frac{\prod_{1 \le \ell < m \le {n+1}} [(w_m - w_\ell) (q\, w_\ell - q^{-1} w_m)]}{\prod_{\ell=1}^{n+1} \left[ \prod_{1 \le i \le b_\ell}(w_\ell - z_i) \prod_{b_\ell \le i \le N}(q\, w_\ell - q^{-1} z_i) \right]},
\end{multline*}
and the integration contours are the same. That Eq.~(\ref{e:b}) and Eq.~(\ref{e:a}) give the same set of functions will be derived below, see Proposition \ref{p:2}.

Using the reasonings similar to those of the paper \cite{DiFZin07}, one can show that $\Psi_{a_1, \ldots, a_n}(z_1,\ldots, z_N)$ and $\overline{\Psi}_{b_1, \ldots, b_{n+1}}(z_1,\ldots, z_N)$ are homogeneous polynomials in the variables $z_1$, $\ldots$, $z_N$ of degree $n(n+1)$.

Let us define two vectors in the subspace of $(\bbC^2)^{\otimes N}$ spanned by the basis vectors with $n$ down arrows,
\[
\Psi(z_1, \ldots, z_N) = \sum_{1 \le a_1 < \ldots < a_n \le N} \Psi_{a_1, \ldots, a_n}(z_1, \ldots, z_N) e_{a_1, \ldots, a_n},
\]
and
\[
\overline{\Psi}(z_1, \ldots, z_N) = \sum_{1 \le b_1 < \ldots < b_{n+1} \le N} \overline{\Psi}_{b_1, \ldots, b_{n+1}}(z_1, \ldots, z_N) \overline{e}_{b_1, \ldots, b_{n+1}},
\]
where $\overline{e}_{b_1, \ldots, b_{n+1}}$ is a basis vector with up arrows in positions $b_1$, $\ldots$, $b_{n+1}$.

We will denote by $\overline{a_1, \ldots, a_n}$ and $\overline{b_1, \ldots, b_{n+1}}$ the increasing sequences of indices complementing the sets $a_1, \ldots, a_n$ and $b_1, \ldots, b_{n+1}$ to the sequence $1, \ldots, 2n+1$. Using this notation, we can write, in particular,
\[
\overline{e}_{\overline{a_1, \ldots, a_n}} = e_{a_1, \ldots, a_n}, \qquad e_{\overline{b_1, \ldots, b_{n+1}}} = \overline{e}_{b_1, \ldots, b_{n+1}}.
\]

\begin{proposition}
For each $i = 1, \ldots, N-1$ the vector $\Psi(z_1, \ldots, z_N)$ satisfies the equation
\begin{equation}
\check R_{i,i+1}(z_{i+1}/z_i) \Psi(z_1, \ldots, z_i, z_{i+1}, \ldots, z_N) = \Psi(z_1, \ldots, z_{i+1}, z_i, \ldots, z_N), \label{e:exch1}
\end{equation}
where $\check R_{i,i+1}(x) = P_{i, i+1} R_{i, i+1}(x)$. The vector $\overline{\Psi}(z_1, \ldots, z_N)$ satisfies the same equations.
\end{proposition}

\begin{proof}
First consider the vector $\Psi(z_1, \ldots, z_{2n+1})$. Let us rewrite explicitly Eq.~(\ref{e:exch1}) in components. There are four cases, depending on the values of spins at locations $i$ and $i+1$. In terms of the integers $a_1$, $\ldots$, $a_n$, it corresponds to:
(i) $i$ and $i+1$ are not among the integers $a_1$, $\ldots$, $a_n$; (ii) for some $\ell < n$ one has $a_\ell = i$ and $a_{\ell+1} \ne i+1$; (iii) for some $l > 1$ one has $a_\ell = i+1$ and $a_{\ell-1} \ne i$; (iv) for some $l < n$ one has $a_\ell = i$ and $a_{\ell+1} = i+1$.

In cases (i) and (iv), Eq.~(\ref{e:exch1}) means that the component $\Psi_{a_1, \ldots, a_n}(z_1, \ldots, z_{2n+1})$ divided by $q \, z_i - q^{-1} z_{i+1}$ must be symmetric in the variables $z_i$, $z_{i+1}$. In case (i) the integrand $\Phi_{a_1, \ldots, a_n}(w_1, \ldots, w_n | z_1, \ldots, z_{2n+1})$ of Eq.~(\ref{e:b}) is symmetric in the exchange of $z_i$ and $z_{i+1}$. This implies that $\Psi_{a_1, \ldots, a_n}(z_1, \ldots, z_{2n+1})$ divided by $q \, z_i - q^{-1} z_{i+1}$ is symmetric in the variables $z_i$, $z_{i+1}$. In the case (iv) $\Phi_{a_1, \ldots, a_n}(\ldots, z_i, z_{i+1}, \ldots) - \Phi_{a_1, \ldots, a_n}(\ldots, z_{i+1}, z_i, \ldots)$ is an antisymmetric function of $w_\ell$ and $w_{\ell+1}$, and integration over these variables gives zero. Therefore, $\Psi_{a_1, \ldots, a_n}(z_1, \ldots, z_{2n+1})$ divided by $q \, z_i - q^{-1} z_{i+1}$ is symmetric in the exchange of $z_i$ and $z_{i+1}$.

In case (ii) we have the non-trivial equation:
\begin{multline}
(q - q^{-1}) z_{i+1} \Psi_{\ldots, i, \ldots}(\{z\}) + (z_{i+1} - z_i) \Psi_{\ldots, i+1, \ldots}(\{z\}) \\ = (q \,z_i-q^{-1}z_{i+1}) \tau_i \Psi_{\ldots, i, \ldots}(\{z\}),
\label{e:ntb}
\end{multline}
where we use the convenient notation $\tau_i$ for the exchange of variables $z_i$ and $z_{i+1}$, and dots stand for the unchanged indices. It is easy to see that in the case under consideration one has
\begin{gather*}
\Phi_{\ldots, i+1, \ldots}(\{w\} | \{z\}) = \frac{z_{i+1}}{z_i} \frac{q \, w_\ell - q^{-1} z_i}{w_\ell - z_{i+1}} \Phi_{\ldots, i, \ldots}(\{w\} | \{z\}), \\
\tau_i \Phi_{\ldots, i, \ldots}(\{w\} | \{z\}) = \frac{q \, z_{i+1}- q^{-1} z_i}{q \, z_i - q^{-1} z_{i+1}} \frac{z_{i+1}}{z_i} \frac{w_\ell - z_i}{w_\ell - z_{i+1}} \Phi_{\ldots, i, \ldots}(\{w\} | \{z\}).
\end{gather*}
Using this relations one can prove that
\begin{multline*}
(q - q^{-1}) z_{i+1} \Phi_{\ldots, i, \ldots}(\{w\} | \{z\}) + (z_{i+1} - z_i) \Phi_{\ldots, i+1, \ldots}(\{w\} | \{z\}) \\ = (q \, z_{i+1} - q^{-1} z_i) \tau_i \Phi_{\ldots, i, \ldots}(\{w\} | \{z\}).
\end{multline*}
Now the validity of Eq.~(\ref{e:ntb}) is evident.

A similar discussion can be made for case (iii) and for the vector $\overline{\Psi}(z_1, \ldots, z_{2n+1})$. Let us only mention that for the non-trivial case analogous to (iii) one should prove the relation
\begin{multline}
(q - q^{-1}) z_{i+1} \overline{\Psi}_{\ldots, i + 1, \ldots}(\{z\}) + (z_{i+1} - z_i) \overline{\Psi}_{\ldots, i, \ldots}(\{z\}) \\ = (q \,z_i-q^{-1}z_{i+1}) \tau_i \overline{\Psi}_{\ldots, i + 1, \ldots}(\{z\}),
\label{e:nta}
\end{multline}
which actually follows from the equalities
\begin{gather*}
\overline{\Phi}_{\ldots, i+1, \ldots}(\{w\} | \{z\}) = \frac{q \, w_\ell - q^{-1} z_i}{w_\ell - z_{i+1}} \, \overline{\Phi}_{\ldots, i+1, \ldots}(\{w\} | \{z\}), \\
\tau_i \overline{\Phi}_{\ldots, i+1, \ldots}(\{w\} | \{z\}) = \frac{q \, z_{i+1}- q^{-1} z_i}{q \, z_i - q^{-1} z_{i+1}} \frac{q \, w_\ell - q^{-1} z_{i+1}}{w_\ell - z_{i+1}} \, \overline{\Phi}_{\ldots, i, \ldots}(\{w\} | \{z\}).
\end{gather*}
Thus, Eq.~(\ref{e:exch1}) is satisfied by both $\Psi(z_1, \ldots, z_N)$ and $\overline{\Psi}(z_1, \ldots, z_N)$.
\end{proof}

\begin{proposition} \label{p:2}
The vectors $\Psi(z_1, \ldots, z_N)$ and $\overline{\Psi}(z_1, \ldots, z_N)$ coincide.
\end{proposition}

\begin{proof}
It is clear from the definition of the vectors $\Psi(z_1, \ldots, z_N)$ and $\overline{\Psi}(z_1, \ldots, z_N)$ that one should prove the equality
\begin{equation}
\Psi_{a_1, \ldots, a_n}(z_1, \ldots, z_N) = \overline \Psi_{\overline{a_1, \ldots, a_n}}(z_1, \ldots, z_N) \label{e:cc}.
\end{equation}
Let us prove this equality for the components $\Psi_{1, \ldots, n}(z_1, \ldots, z_N)$ and $\overline{\Psi}_{\overline{1, \ldots, n}}(z_1, \ldots, z_N) = \overline{\Psi}_{n+1, \ldots, N}(z_1, \ldots, z_N)$. Using Eq.~(\ref{e:b}), we see that the only poles that contribute to the integral for $\Psi_{1, \ldots, n}(z_1, \ldots, z_N)$ are the poles at the points $w_i = z_i$, $i = 1, \ldots, n$, and we find immediately
\begin{equation}
\Psi_{1,\ldots,n}(z_1, \ldots, z_N) = \prod_{i=1}^n z_i \prod_{1\le i < j \le n}(q \, z_i - q^{-1}z_j) \prod_{n + 1 \le i < j \le N}(q \, z_i - q^{-1} z_j).
\label{e:base}
\end{equation}
Consider now Eq.~(\ref{e:a}) for $\Psi_{n+1,\ldots,N}(z_1, \ldots, z_N)$. We note that one can pick up poles outside the contours instead of inside, and that there are none at infinity. Hence, the only poles that contribute to the integral are the poles at the points $w_i = z_{n+i}$, $i=1,\ldots,n+1$, and the computation leads to the same expression (\ref{e:base}). Thus, we see that $\Psi_{1, \ldots, n}(z_1, \ldots, z_N) = \overline{\Psi}_{n+1, \ldots, N}(z_1, \ldots, z_N)$.

It is easy to see that any component of $\Psi(z_1, \ldots, z_N)$ can be obtained from the component $\Psi_{1, \ldots, n}(z_1, \ldots, z_N)$ by incrementing indices with the help of the equality
\[
\Psi_{\ldots, i+1, \ldots}(\{z\}) \\= \frac{q \, z_i - q^{-1} z_{i+1}}{z_{i+1} - z_i} \tau_i \Psi_{\ldots, i, \ldots}(\{z\}) - (q - q^{-1}) {\frac{z_{i+1}}{z_{i+1} - z_i}} \Psi_{\ldots, i, \ldots}(\{z\}),
\]
which follows from Eq.~(\ref{e:ntb}). Similarly, any component of $\overline{\Psi}(z_1, \ldots, z_N)$ can be obtained from the component $\overline{\Psi}_{n+1, \ldots, N}(z_1, \ldots, z_N)$ by decrementing indices with the help of the equality
\[
\overline{\Psi}_{\ldots, i, \ldots}(\{z\}) = \frac{q \,z_i-q^{-1}z_{i+1}}{z_{i+1} - z_i} \tau_i \overline{\Psi}_{\ldots, i + 1, \ldots}(\{z\}) - (q - q^{-1}) \frac{z_{i+1}}{z_{i+1} - z_i} \overline{\Psi}_{\ldots, i + 1, \ldots}(\{z\}),
\]
which follows from Eq.~(\ref{e:nta}). Comparing now the two above equalities, we conclude that Eq.~(\ref{e:cc}) is valid for all components.
\end{proof}

Let $\sigma$ be the operator of left rotation in the space $(\bbC^2)^{\otimes N}$ defined by the equality
\[
\sigma (v_1 \otimes v_2 \otimes \cdots \otimes v_N) = v_2 \otimes \cdots \otimes v_N \otimes v_1.
\]

\begin{proposition}
The vector $\Psi(z_1, \ldots, z_N)$ satisfies the following cyclicity condition
\begin{equation}
D_N \, \sigma \, \Psi(z_1, \ldots, z_N) = \Psi(z_2, \ldots, z_N, s z_1), \label{e:rota}
\end{equation}
where $s = q^6$ and $D_N$ is the operator
\[
D = q^{3n} q^{3(\sigma^z + 1)/2}
\]
acting in the last factor of the tensor product $(\bbC^2)^{\otimes N}$.
\end{proposition}

\begin{proof}
First of all write Eq.~(\ref{e:rota}) in terms of components. We have two cases in terms of the components $\Psi_{a_1, \ldots, a_n}$:
\begin{multline}
\Psi_{a_1, \ldots, a_{n-1}, a_n}(z_2, \ldots, z_N, s z_1) \\
= \begin{cases}
q^{3n+3} \, \Psi_{1, a_1+1, \ldots, a_{n-1}+1}(z_1, z_2, \ldots, z_N), \quad & a_n = N; \quad \\
q^{3n} \, \Psi_{a_1+1, \ldots, a_{n-1}+1, a_n+1}(z_1, z_2, \ldots, z_N), & a_n < N.
\end{cases} \label{e:cs1}
\end{multline}
and two similar cases in terms of the components $\overline{\Psi}_{b_1, \ldots, b_{n+1}}$:
\begin{multline}
\overline{\Psi}_{b_1, \ldots, b_n, b_{n+1}}(z_2, \ldots, z_N, s z_1) \\
= \begin{cases}
q^{3n+3} \, \overline{\Psi}_{b_1+1, \ldots, b_n+1, b_{n+1}+1}(z_1, z_2, \ldots, z_N), \quad & b_{n+1} < N; \quad \\
q^{3n} \, \overline{\Psi}_{1, b_1+1, \ldots, b_n+1}(z_1, z_2, \ldots, z_N), & b_{n+1} = N.
\end{cases} \label{e:cs2}
\end{multline}
Taking into account the correspondence described by Eq.~(\ref{e:cc}),
it is easy to convince oneself that the validity of the first case of Eq.~(\ref{e:cs1}) implies the validity of the first case of Eq.~(\ref{e:cs2}), the validity of the second case of Eq.~(\ref{e:cs1}) implies the validity of the second case of Eq.~(\ref{e:cs2}).

Let us start with the first case of Eq.~(\ref{e:cs1}). Picking up the poles in Eq.~(\ref{e:b}) outside the contours, we see that there is a single pole that contributes to the integral for $\Psi_{a_1, \ldots, a_{n-1}, N}(z_1, \ldots, z_N)$ over $w_n$, namely, the pole at the point $w_n = q^{-2} s\, z_1 = q^4 z_1$. Using this fact, we find that
\begin{multline*}
\Psi_{a_1, \ldots, a_{n-1}, N}(z_2, \ldots, z_{2n+1}, s z_1) \\
= q^{3n+3} (q - q^{-1})^{n-1} z_1 \prod_{\ell=1}^{n-1} z_{a_\ell+1} \prod_{2 \le i < j \le N}(q\,z_i-q^{-1}z_j) \\
\times \oint \cdots \oint \prod_{\ell=1}^{n-1} \frac{w_\ell \rmd w_\ell}{2 \pi \rmi}
\frac{\prod_{\ell=1}^{n-1}(q \, z_1-q^{-1}w_\ell)\prod_{1 \le \ell < m \le n-1} [(w_m - w_\ell) (q\, w_\ell - q^{-1} w_m)]} {\prod_{\ell=1}^{n-1} \left[ \prod_{1 \le i \le a_\ell}(w_\ell - z_{i+1}) \prod_{a_\ell \le i \le 2n}(q\, w_\ell - q^{-1} z_{i+1}) \right]}.
\end{multline*}
On the other hand, there is a single pole contributing to the integral for $\Psi_{1, a_1+1, \ldots, a_{n-1}+1}(z_1, \ldots, z_N)$ over $w_1$, namely, the pole at the point $w_1=z_1$, and we obtain
\begin{multline*}
\Psi_{1, a_1+1, \ldots, a_{n-1}+1}(z_1, \ldots, z_N) \\
= (q - q^{-1})^{n-1} z_1\prod_{\ell=1}^{n-1} z_{a_\ell+1}
\prod_{2 \le i < j \le N}(q \,z_i-q^{-1}z_j) \\
\times \oint \cdots \oint \prod_{\ell=2}^{n} \frac{w_\ell \rmd w_\ell}{2 \pi \rmi}
\frac{\prod_{\ell=2}^n(q \,  z_1-q^{-1}w_\ell)
\prod_{2 \le \ell < m \le n} [(w_m - w_\ell) (q\, w_\ell - q^{-1} w_m)]}
{
\prod_{\ell=2}^n \left[ \prod_{2 \le i \le a_{\ell-1}+1}(w_\ell - z_i) \prod_{a_{\ell-1}+1 \le i \le 2n + 1}(q\, w_\ell - q^{-1} z_i) \right]}.
\end{multline*}
After the evident change of integration variables one sees that the first case of Eq.~(\ref{e:cs1}) is true. In the same way one can show that the second case of Eq.~(\ref{e:cs2}) is also true. As is noted above it suffices to prove the validity of all cases of Eqs. (\ref{e:cs1}) and (\ref{e:cs2}).
\end{proof}

\begin{example}
For $n = 1$, that is in size $N=3$, from either expression (\ref{e:b}) or (\ref{e:a}) we obtain
\begin{align*}
\Psi_{1}(z_1, z_2, z_3) &= \overline{\Psi}_{2, 3}(z_1, z_2, z_3) = z_1 (q \, z_2 - q^{-1} z_3), \\[.4em]
\Psi_{2}(z_1, z_2, z_3) &= \overline{\Psi}_{1, 3}(z_1, z_2, z_3) = z_2 (q^{-2} z_3 - q^2 z_1), \\[.4em]
\Psi_{3}(z_1, z_2, z_3) &= \overline{\Psi}_{1, 2}(z_1, z_2, z_3) =  z_3(q \, z_1 - q^{-1} z_2).
\end{align*}
One can easily see that these expressions satisfy Eqs. (\ref{e:cs1}) and (\ref{e:cs2}).
\end{example}

Eqs.~(\ref{e:exch1}) and (\ref{e:rota}) are the $q$KZ equations in the form proposed by Smirnov \cite{Smi86}.

\subsection{From $q$KZ to the six-vertex model and XXZ spin chain}

Now along the lines of the paper \cite{KasPas06} we prove the following proposition.

\begin{proposition}
When $q = \rme^{\pm 2\pi\rmi/3}$ the vector $\Psi(z_1, \ldots, z_N)$ is an eigenvector of the transfer matrix $T(x | z_1, \ldots, z_N)$ with the eigenvalue $1$.
\end{proposition}

\begin{proof}
Let us start with the vector $\Psi(z_i, z_1, \ldots, z_{i-1}, \hat z_i, z_{i+1}, \ldots, z_N)$. Acting on it successively by the operator $\check R_{1,2}(z_1/z_i)$, \ldots, $\check R_{i-1,i}(z_{i-1}/z_i)$ and having in mind Eq. (\ref{e:exch1}), we obtain
\begin{multline}
\check R_{i-1,i}(z_{i-1}/z_i) \ldots \check R_{1,2}(z_1/z_i)\Psi(z_i, z_1, \ldots, z_{i-1}, \hat z_i, z_{i+1}, \ldots, z_N) \\
= \Psi(z_1, z_2, \ldots, z_{i-1}, z_i, z_{i+1}, \ldots, z_N). \label{e:a3}
\end{multline}
Now consider the vector $D_N \, \sigma \, \Psi(z_i, z_1, \ldots, z_{i-1}, \hat z_i, z_{i+1}, \ldots, z_N)$. It follows from Eq. (\ref{e:rota}) that
\[
D_N \,  \sigma \, \Psi(z_i, z_1, \ldots, z_{i-1}, \hat z_i, z_{i+1}, \ldots, z_N) = \Psi(z_1, \ldots, z_{i-1}, \hat z_i, z_{i+1}, \ldots, z_N, s z_i).
\]
The successive action of the operators $\check R_{N-1,N}(s z_i/z_N)$, \ldots, $\check R_{i,i+1}(s z_i/z_{i+1})$ gives
\begin{multline}
\check R_{i,i+1}(s z_i/z_{i+1}) \ldots \check R_{N-1,N}(s z_i/z_N) \\
\times D_N \, \sigma \, \Psi(z_i, z_1, \ldots, z_{i-1}, \hat z_i, z_{i+1}, \ldots, z_N) \\
= \Psi(z_1, z_2, \ldots, z_{i-1}, s z_i, z_{i+1}, \ldots, z_N). \label{e:a4}
\end{multline}
Using the equality $\check R(x) \check R(1/x) = 1$, rewrite Eq. (\ref{e:a3}) as
\begin{multline*}
\Psi(z_i, z_1, \ldots, z_{i-1}, \hat z_i, z_{i+1}, \ldots, z_N) \\
= \check R_{1,2}(z_i/z_1) \ldots \check R_{i-1,i}(z_i/z_{i-1}) \Psi(z_1, z_2, \ldots, z_{i-1}, z_i, z_{i+1}, \ldots, z_N).
\end{multline*}
Substituting this equality into Eq. (\ref{e:a4}), we obtain
\begin{equation}
S_i(z_1, \ldots, z_N) \Psi(z_1, \ldots, z_N),
= \Psi(z_1, \ldots, s z_i, \ldots z_N),
\label{e:realqkz}
\end{equation}
where $S_i(z_1, \ldots, z_N)$ are the Yang scattering matrices,
\begin{multline*}
S_i(z_1, \ldots, z_N) = \check R_{i,i+1}(s z_i/z_{i+1}) \ldots \check R_{N-1,N}(s z_i/z_N)\\
\times D_N \, \sigma \, \check R_{1,2}(z_i/z_1) \ldots \check R_{i-1,i}(z_i/z_{i-1}).
\end{multline*}

One can show that when $q = \rme^{\pm 2 \pi \rmi/3}$ the Yang scattering matrices are related to the transfer matrix by the equality $S_i(z_1, \ldots, z_N) = T(z_i | z_1, \ldots, z_N)$. Therefore, in this case
\begin{equation}
T(z_i | z_1, \ldots, z_N) \Psi(z_1, \ldots, z_N) = \Psi(z_1, \ldots, z_N). \label{e:ti}
\end{equation}

Let us demonstrate now that when $q = \rme^{\pm 2\pi\rmi/3}$ the vector $\Psi(z_1, \ldots, z_N)$ is an eigenvector of $T(0 | z_1, \ldots, z_N)$ with the eigenvalue $1$. To this end define the $2 \times 2$ matrices $\calR_i(x)$, $i = 1, \ldots, N$, with the matrix elements being operators in the space $(\bbC^2)^{\otimes N}$ as
\[
\calR_i(x) = \left( \begin{array}{cc}
\displaystyle \frac{a(x) + b(x)}{2} + \frac{a(x) - b(x)}{2} \sigma^z_i & c(x) \sigma^-_i \\
c'(x) \sigma^+_i & \displaystyle \frac{a(x) + b(x)}{2} - \frac{a(x) - b(x)}{2} \sigma^z_i
\end{array} \right).
\]
One can easily show that
\[
T(x | z_1, \ldots, z_N) = \mathrm{tr} [\calR_1(x/z_1) \ldots \calR_N(x/z_N)].
\]
At $x = 0$ one obtains
\[
\calR_i(0) = - q^{-2} \left( \begin{array}{cc}
\displaystyle \frac{1 + \sigma^z_i}{2} + q \frac{1 - \sigma^z_i}{2} & 0 \\
- q (q - q^{-1}) & \displaystyle \frac{1 - \sigma^z_i}{2} + q \frac{1 + \sigma^z_i}{2}
\end{array} \right),
\]
therefore,
\begin{multline*}
T(0 | z_1, \ldots, z_N) = (-q^{-2})^N  \left( \frac{1 + \sigma^z_1}{2} + q \frac{1 - \sigma^z_1}{2} \right) \ldots \left( \frac{1 + \sigma^z_N}{2} + q \frac{1 - \sigma^z_N}{2} \right) \\
+ (-q^{-2})^N  \left( \frac{1 - \sigma^z_1}{2} + q \frac{1 + \sigma^z_1}{2} \right) \ldots \left( \frac{1 - \sigma^z_N}{2} + q \frac{1 + \sigma^z_N}{2} \right).
\end{multline*}
This equality implies that any vector belonging to the sector with $K$ down arrows is an eigenvector of $T(0 | z_1, \ldots, z_N)$ with the eigenvalue
\[
\lambda = \lambda(0 | z_1, \ldots, z_N) = (-1)^N q^{-2N} q^K + (-1)^N q^{-2N} q^{N-K} = (-1)^N (q^{-2N+K} + q^{-N-K}).
\]
One can convince oneself  that in the case when $N = 2n+1$, $K = n$,
and $q = \rme^{\pm 2\pi\rmi/3}$ one has $\lambda = 1$. Therefore, in this case
\begin{equation}
T(0 | z_1, \ldots, z_N) \Psi(z_1, \ldots, z_N) =  \Psi(z_1, \ldots, z_N). \label{e:t0}
\end{equation}

From the definition of the transfer matrix $T(x | z_1, \ldots, z_N)$ it follows that the vector $\prod_{i=1}^N(q - q^{-1} x z_i^{-1}) T(x | z_1, \ldots, z_N) \Psi(z_1, \ldots, z_N)$ is a polynomial in $x$ of degree $N$ with coefficients in $(\bbC^2)^{\otimes N}$. Eqs. (\ref{e:ti}) and (\ref{e:t0}) say that this vector coincide with $\prod_{i=1}^N(q - q^{-1} x z_i^{-1}) \Psi(z_1, \ldots, z_N)$ for $N + 1$ values of $x$. Hence, it coincides with $\prod_{i=1}^N(q - q^{-1} x z_i^{-1}) \Psi(z_1, \ldots, z_N)$ for any value of $x$. Thus, the vector $\Psi(z_1, \ldots, z_N)$ is an eigenvector of the transfer matrix $T(x | z_1, \ldots, z_N)$ with the eigenvalue $1$.
\end{proof}

In the homogeneous limit for general $q$ the logarithmic derivative of
the transfer matrix $T(x) = T(x | 1, \ldots,1)$ at $x = 1$ is related
to the Hamiltonian $H_{\mathrm{XXZ}}(\Delta)$ of the XXZ spin model
via Eq.~(\ref{e:ld}). Therefore, in the case when $q = \rme^{\pm 2\pi\rmi/3}$ the vector $\Psi(1, \ldots, 1)$ is an eigenvector of $H_{\mathrm{XXZ}}(-1/2)$ with the eigenvalue $-3N/4$. There is in fact a more direct derivation of the latter.

Indeed, starting from Eq. (\ref{e:ti}) for any distinct $i$ and $j$ one writes
\begin{equation}
S_i^{}(z_1, \ldots, z_N) S_j^{-1}(z_1, \ldots, z_N) \Psi(z_1, \ldots, z_N) = \Psi(z_1, \ldots, z_N). \label{e:evs}
\end{equation}
Now for infinitesimal $\epsilon$ and $\epsilon'$ let us set $z_i = 1 + \epsilon$, $z_j = 1 + \epsilon'$, $z_k = 1$ for $k \ne i, j$. From the definition of $S_i(z_1, \ldots, z_N)$ it follows that in this case for $i < j$ one has
\[
S_i  = \left( 1 + \epsilon \sum_{k=i}^{N-1} h_k - \epsilon' h_{j-1} + \epsilon \sum_{k=1}^{i-1} \sigma h_k \sigma^{-1} + \cdots \right) \sigma,
\]
where $h_k = \check R'_k(1)$. Using the equality $\sigma \check R_i(x) \sigma^{-1} = \check R_{i-1}(x)$ and introducing the missing operator $\check R_N(x)$ which acts on spaces $1$ and $N$, such that $\check R_N(x) = \sigma \check R_1(x) \sigma^{-1}$, we come to the equality
\[
S_i = \left( 1 + \epsilon \sum_{\substack{k=1 \\ k \ne i-1}}^{N} h_k - \epsilon' h_{j-1} + \cdots \right) \sigma.
\]
In a similar way one obtains
\[
S_j = \left( 1 + \epsilon' \sum_{\substack{k=1 \\ k \ne j-1}}^{N} h_k - \epsilon h_{i-1} + \cdots \right) \sigma,
\]
so that
\[
S_i^{} S_j^{-1}=1+(\epsilon-\epsilon')\sum_{k=1}^N h_k +\cdots.
\]
In the case $i > j$ we come to the same result.

Expanding Eq.~(\ref{e:evs}) to first order in $\epsilon$ and $\epsilon'$, we conclude that the vector $\Psi(1, \ldots, 1)$ is an eigenvector of $\sum_{i=1}^N h_i$ with the eigenvalue $0$. One can show that $\sum_{i=1}^N h_i$ coincides with the logarithmic derivative of the transfer matrix $T(x) = T(x | 1, \ldots,1)$ at $x = 1$, which, in turn, is related to the Hamiltonian $H_{\mathrm{XXZ}}(\Delta)$ of the XXZ spin chain via Eq.~(\ref{e:ld}). Thus, we again see that the vector $\Psi(1, \ldots, 1)$ is an eigenvector of the Hamiltonian $H_{\mathrm{XXZ}}(-1/2)$ with the eigenvalue $-3 N/4$.

Concluding this section, note that using the identity
\[
\sigma = P_{N-1,N} \ldots P_{2,3} P_{1,2},
\]
one can rewrite the definition of the Yang scattering matrices as
\begin{multline*}
S_i(z_1, \ldots, z_N) = R_{i,i+1}^{-1}(s^{-1} z_{i+1}/z_i) \ldots R_{i,N}^{-1}(s^{-1} z_N/z_i)\\
\times D_i \, R_{1,i}(z_i/z_1) \ldots R_{i-1,i}(z_i/z_{i-1}).
\end{multline*}
Then Eq.~(\ref{e:realqkz}) takes the form
\begin{multline*}
\Psi(z_1, \ldots, s z_i, \ldots z_N) = R_{i,i+1}^{-1}(s^{-1} z_{i+1}/z_i) \ldots R_{i,N}^{-1}(s^{-1} z_N/z_i) \\
\times D_i \, R_{1,i}(z_i/z_1) \ldots R_{i-1,i}(z_i/z_{i-1}) \Psi(z_1, \ldots, z_i, \ldots z_N).
\end{multline*}
It is this equation that is now usually called the $q$KZ equation, see, for example, the paper \cite{FreRes92}.

\section{Homogeneous limit and applications} \label{s:5}

We will now find the homogeneous limit of the representations (\ref{e:b}) and (\ref{e:a}) for the components of the vector $\Psi(z_1, \ldots, z_N)$. In this limit $\Psi(z_1, \ldots, z_N)$ depends on the single remaining parameter $q$, and when $q = \rme^{\pm 2\pi\rmi/3}$ it coincides with the ground state of the Hamiltonian $H_{\mathrm{XXZ}}(-1/2)$. Thus, the homogeneous limit of the representations~(\ref{e:b}) and~(\ref{e:a}) at $q = \rme^{\pm 2\pi\rmi/3}$ give the components of the ground state vector of the XXZ spin chain at $\Delta = -1/2$.

\subsection{Integral formulae in the homogeneous limit}

It is convenient to define
\begin{multline*}
\psi = (q - q^{-1})^{-n^2} \Psi(1, \ldots, 1) \\
= \sum_{1 \le a_1 < \ldots < a_n \le N}
\psi_{a_1, \ldots, a_n} e_{a_1, \ldots, a_n} = \sum_{1 \le b_1 <
  \ldots < b_{n+1} \le N} \overline{\psi}_{b_1, \ldots, b_{n+1}}
\overline{e}_{b_1, \ldots, b_{n+1}}.
\end{multline*}
The normalization is chosen in such a way that $\psi_{1,2,\ldots,n} = \overline{\psi}_{1, 2, \ldots, n+1} = 1$.

After the change of variables
\[
u_\ell = \frac{w_\ell - 1}{q w_\ell - q^{-1}}
\]
we obtain from Eq.~(\ref{e:b})
\begin{equation}
\psi_{a_1, \ldots, a_n} = \oint \ldots \oint \prod_{\ell=1}^n \frac{\rmd u_\ell (1 + \tau u_\ell + u_\ell^2)}{2 \pi \rmi \,u_\ell^{a_\ell}} \prod_{1 \le \ell < m \le n} [(u_m - u_\ell) (1 + \tau u_m + u_\ell u_m)], \label{e:hl1}
\end{equation}
where $\tau = - q - q^{-1}$. In a similar way Eq.~(\ref{e:a}) yields
\begin{equation}
\overline{\psi}_{b_1, \ldots, b_{n+1}} =
\oint\cdots\oint \prod_{\ell=1}^{n+1} \frac{\rmd u_\ell}{2\pi \rmi \, u_\ell^{b_\ell}}
\prod_{1\le \ell<m\le n+1} [(u_m-u_\ell)(1 + \tau u_m + u_\ell u_m)]. \label{e:hl2}
\end{equation}
It is not difficult to show that we really have
\[
\psi_{1, 2, \ldots, n} = 1, \qquad \overline{\psi}_{1, 2, \ldots, n+1} = 1.
\]
Recall that the two types of components are related by
\begin{equation}
\psi_{a_1, \ldots, a_n} = \overline{\psi}_{\overline{a_1, \ldots, a_n}}. \label{e:ce}
\end{equation}

After the change of the variables $u_\ell \mapsto 1/u_\ell$ one finds the following reflection symmetry:
\[
\psi_{a_1, \ldots, a_n} = \psi_{N + 1 - a_n, \ldots, N + 1 - a_1}, \qquad \overline{\psi}_{b_1, \ldots, b_{n+1}} = \overline{\psi}_{N + 1 - b_{n+1}, \ldots, N + 1 - b_1}.
\]

\begin{example} Here are all the homogeneous components for $n=2$:
\begin{gather}
\psi_{1,2} =  \psi_{4,5} = 1, \qquad \psi_{1,5} =  \tau, \qquad \psi_{2,3} =  \psi_{3,4} = \tau^2,
\notag \\[-.6em] \label{e:hc} \\[-.6em]
\psi_{1,3} =  \psi_{3,5} = 2 \tau, \qquad \psi_{2,4} =  \tau(1 + \tau^2),
\qquad \psi_{1,4} = \psi_{2,5}= 1 + \tau^2. \notag
\end{gather}
The case $q = \rme^{\pm 2\pi\rmi/3}$, when the vector $\psi$
becomes the ground state of the XXZ spin chain, corresponds to $\tau=1$.
In this case, all components on either line of Eq.~(\ref{e:hc}) become equal. This is a sign of the rotational invariance
\[
\sigma \psi = \psi
\]
which is a direct consequence of Eq.~(\ref{e:rota}).
\end{example}

Note that Eq.~(\ref{e:hl1}) and Eq.~(\ref{e:hl2}) express components of $\psi$ as particular coefficients of a polynomial in the variables $u_\ell$ and $\tau$ with integer coefficients.
An important consequence is that the components of $\psi$ are polynomials in $\tau$ with {\em integer} coefficients. In particular, at $\tau=1$, the components of the ground state of the XXZ spin chain are integers. Thus, Theorem \ref{c:1} is proved. Recall that the normalization of $\psi$ is fixed by $\psi_{1,\ldots,n} = 1$, so that this component is necessarily the smallest one.

\subsection{Recurrence relation}

When $a_1 = 1$ or $b_1 = 1$ it is easy to integrate over $u_1$ in Eq.~(\ref{e:hl1}) or Eq.~(\ref{e:hl2}), one just needs to set $u_1 = 0$. For example, Eq.~(\ref{e:hl1}) gives
\begin{equation*}
\psi_{1, a_2, \ldots, a_n}
= \oint\cdots\oint \prod_{\ell=2}^n \frac{\rmd u_\ell (1 + u_\ell + u_\ell^2) (1 +\tau u_\ell)}{ 2\pi \rmi \, u_\ell^{a_\ell - 1}}
\prod_{2 \le \ell < m \le n} [(u_m-u_\ell)(1 + \tau u_m + u_\ell u_m)].
\end{equation*}
That implies a recurrence relation
\begin{equation}
\psi^{(n)}_{1, a_2, \ldots, a_n} =
\sum_{\varepsilon_2, \ldots, \varepsilon_n \in \{0,1\}} \tau^{\sum_{\ell=2}^n\varepsilon_\ell}
\psi^{(n-1)}_{a_2 -1 -\varepsilon_2,\ldots, a_n -1 - \varepsilon_n}, \label{e:rra}
\end{equation}
where for clarity we use additional indices to stress that the components in the left and right hand sides correspond to different values of $n$. The obtained recurrence relation is not a closed recurrence because some indices may become equal in the right hand side of the formula.
A similar relation is satisfied by the components $\overline{\psi}_{b_1, \ldots, b_{n+1}}$.

We now assume that $q = \rme^{\pm 2\pi\rmi/3}$, that is $\tau=1$. Using rotational invariance of $\psi$, one can rewrite Eq.~(\ref{e:rra}) in the form
\begin{equation}
\psi^{(n)}_{a_1, a_2, \ldots, a_n} =
\sum_{\varepsilon_2, \ldots, \varepsilon_n \in \{0,1\}}
\psi^{(n-1)}_{a_2 -a_1 -\varepsilon_2,\ldots, a_n -a_1 - \varepsilon_n}, \label{e:rr}
\end{equation}

\subsection{Proof of second theorem}

Applying Eq.~(\ref{e:rr}) to the component $\psi^{(n)}_{1, 3, \ldots, 2n - 1}$, after the change  $\varepsilon_\ell \mapsto 1 - \varepsilon_{\ell - 1}$ one obtains
\begin{equation}
\psi^{(n)}_{1, 3, \ldots, 2n - 1} = \sum_{\varepsilon_1, \ldots, \varepsilon_{n-1} \in \{0,1\}}
\psi^{(n-1)}_{1 + \varepsilon_1, \ldots, 2n - 3 + \varepsilon_{n-1}}. \label{e:ps}
\end{equation}
We then use Eq.~(\ref{e:prerefasmx}), which we rewrite at $n-1$:
\begin{equation}
\frac{1}{\psi^{(n-1)}_{1, 3, \ldots, 2n-3}} \sum_{\varepsilon_1, \ldots, \varepsilon_{n-1} \in \{0,1\}} \alpha^{\sum_{\ell=1}^{n-1} \varepsilon_\ell} \, \psi^{(n-1)}_{1 + \varepsilon_1, 3 + \varepsilon_2, \ldots, 2n-3 + \varepsilon_{n-1}} = \frac{1}{A(n-1)} \sum_{r=1}^n \alpha^{r-1} A(n,r).
\label{e:prerefasm}
\end{equation}
Recall that $A(n, r)$ are the numbers providing the refined enumeration of the alternating sign matrices. At $\alpha = 1$, taking into account that $\sum_{r=1}^{n} A(n,r) = A(n)$, we have
\[
\sum_{\varepsilon_1, \ldots, \varepsilon_{n-1} \in \{0,1\}} \psi^{(n-1)}_{1 + \varepsilon_1, 3 + \varepsilon_2, \ldots, 2n-3 + \varepsilon_{n-1}} = \frac{A(n)}{A(n-1)} \psi^{(n-1)}_{1, 3, \ldots, 2n-3}.
\]
Comparing this equality with Eq.~(\ref{e:ps}), we conclude that
\[
\psi^{(n)}_{1, 3, \ldots, 2n-1} = \frac{A(n)}{A(n-1)} \psi^{(n-1)}_{1, 3, \ldots, 2n-3}.
\]
Now taking into account that $\psi^{(1)}_1 = 1 = A(1)$, we come to the equality
\[
\psi^{(n)}_{1, 3, \ldots, 2n-1} = A(n)
\]
which is the statement of Theorem \ref{c:2}.

One can now simplify the equality (\ref{e:prerefasm}) to
\begin{equation}
\sum_{\varepsilon_1, \ldots, \varepsilon_{n-1} \in \{0,1\}} \alpha^{\sum_{\ell=1}^{n-1}
  \varepsilon_\ell} \, \psi^{(n-1)}_{1 + \varepsilon_1, 3 + \varepsilon_2, \ldots,
  2n-3 + \varepsilon_{n-1}} = \sum_{r=1}^n \alpha^{r-1} A(n,r).
\label{e:refasm}
\end{equation}
which was the main observation of the paper \cite{RazStr06}. This results in the following integral representation for the generating function of the refined enumeration of the alternating sign matrices:
\begin{multline*}
\sum_{r=1}^n \alpha^{r-1} A(n, r) \\
= \oint \cdots \oint \prod_{\ell=1}^{n-1} \frac{\rmd u_\ell (1 + u_\ell + u_\ell^2) (1 + \alpha u_\ell)}{ 2\pi \rmi \, u_\ell^{2\ell}}
\prod_{1 \le \ell<m\le {n-1}} [(u_m-u_\ell)(1 + u_m + u_\ell u_m)].
\end{multline*}

Using the connection between components $\psi^{(n)}_{a_1, \ldots, a_n}$ and $\overline{\psi}^{(n)}_{b_1, \ldots, b_{n+1}}$, and the integral representation (\ref{e:hl2}), one can also prove the validity of the equality
\[
\sum_{r=1}^n \alpha^r A(n, r) = \oint \cdots \oint \prod_{\ell=1}^{n} \frac{\rmd u_\ell (1 + \alpha u_\ell)}{ 2\pi \rmi \, u_\ell^{2\ell}} \prod_{1 \le \ell<m\le {n}} [(u_m-u_\ell)(1 + u_m + u_\ell u_m)].
\]

\subsection{Relation to Temperley--Lieb loop model}

In the paper \cite{DiFZin07}, in the study of $q$KZ solutions in the loop basis,
certain integrals similar to those given by Eq.~(\ref{e:b}) and Eq.~(\ref{e:a}) were introduced. They depend on a sequence of integers, which we shall name
here $b_1,\ldots,b_n$ for consistency with our present notations, and are of the form
\begin{multline*}
\Xi_{b_1, \ldots, b_n}(z_1, \ldots, z_{2n}) = \prod_{1\le i<j\le 2n}(q\,z_i-q^{-1}z_j) \\
\times \oint\cdots \oint
\prod_{\ell=1}^n {\frac{\rmd w_\ell}{2\pi \rmi}}
\frac{
\prod_{1\le \ell<m\le n} [(w_m-w_\ell)(q\,w_\ell-q^{-1}w_m)]}
{\prod_{\ell=1}^n \left[ \prod_{1\le i\le b_\ell} (w_\ell-z_i)
\prod_{b_\ell<i\le 2n} (q\,w_\ell-q^{-1}z_i) \right]
}.
\end{multline*}
Note that these do not exactly coincide with either of Eq.~(\ref{e:b}) and Eq.~(\ref{e:a}),
However, quite remarkably, and as can be easily checked by direct comparison, in the homogeneous limit,
after proper normalization, they are exactly equal to $\overline{\psi}^{(n-1)}_{b_1,\ldots,b_n}$, where once again the size of the system is written explicitly in superscript.

Let us now borrow without proof the following result from the paper \cite{DiFZin07},
see in particular its appendix A. In general, it expresses the aforementioned integrals
as linear combinations of components of the solution of the $q$KZ equation in the loop basis.
These are relations of the form
\begin{equation}
\Xi_{b_1,\ldots,b_n}(z_1, \ldots, z_{2n}) = \sum_{\pi \in \Pi_{2n}} C_{b_1,\ldots,b_n}^\pi \, \Xi_\pi(z_1, \ldots, z_{2n}). \label{e:xicxi}
\end{equation}
The coefficients are given explicitly by the relation
\begin{equation}
C_{b_1,\ldots,b_n}^\pi = \prod_{i<\pi(i)} U_{\# \{ \ell: i\le b_\ell<\pi(i)\} - (\pi(i)-i+1)/2},
\label{e:coeffloop}
\end{equation}
where $i$ and $\pi(i)$ are connected by an arch, and the numbers $U_k$ are defined by the relation
\[
U_{k-1} = \frac{q^{k}-q^{-k}}{q-q^{-1}}.
\]
For example, one has $U_{-1}=0$, $U_0=1$, $U_1=-\tau$, $U_2=\tau^2-1$, etc.
What is relevant here is the homogeneous limit and the value $q = \rme^{\pm 2\pi\rmi/3}$. In this case Eq.~(\ref{e:xicxi}) becomes
\begin{equation}
\bar\psi^{(n-1)}_{b_1,\ldots,b_n}=\sum_{\pi\in\Pi_{2n}} C^\pi_{b_1,\ldots,b_n} \xi_\pi,
\end{equation}
thus relating eigenvectors of XXZ model in size $2n-1$ and of the Temperley--Lieb loop model of
Section~\ref{s:2.3} in size $2n$.

In the paper \cite{RazStr06}, special sequences of the form $a_1=1$ or $2$, \dots, $a_\ell=2\ell-1$ or $2\ell$, \dots, $a_n=2n-1$ or $2n$ were investigated, see Section~\ref{s:3.2}.
Interestingly, something similar can be defined here.
Consider sequences of the form $b_1=1$ or $2$, \dots, $b_\ell=2\ell-1$ or $2\ell$, \dots,
$b_{n-1}=2n-3$ or $2n-2$, $b_n=2n-1$. Note that $b_n=2n$ is not allowed since
in this case $\overline{\psi}_{b_1,\ldots,b_n}$ would be zero.
Applying Eq.~(\ref{e:coeffloop}), we find that the corresponding coefficients $C_{b_1,\ldots,b_n}^\pi$
can only be zero or one, and that they are one if and only if the following condition is met~\cite{DiFZin07}:
\begin{equation}
2\ell<\pi(2\ell)\ \mbox{iff}\ b_\ell=2\ell,
\qquad
\ell=1,\ldots,n-1,
\label{e:parloop}
\end{equation}
or, alternatively,
\begin{equation}
2\ell > \pi(2\ell)\ \mbox{iff}\ b_\ell=2\ell-1,
\qquad
\ell=1, \ldots, n-1.
\end{equation}
Let us call the opening (closing) of $\pi$ an index $i$ such that $i<\pi(i)$ ($\pi(i)<i$),
that is the arch starting at $i$ connects a vertex to the right (left) of $i$. Thus, these particular components $\overline{\psi}_{b_1,\ldots,b_n}$ are partial
sums of $\xi_\pi$ at fixed locations of even openings/closings. In particular, by summing over all such sequences we obtain the full sum of loop components:
\[
\sum_{\varepsilon_1,\ldots,\varepsilon_{n-1}\in\{0,1\}}
\overline{\psi}^{(n-1)}_{2-\varepsilon_1,\ldots,2n-2-\varepsilon_n,2n-1}
=\sum_{\pi\in\Pi_{2n}} \xi_\pi.
\]

The results of the previous section can now be reinterpreted in terms of the Temperley--Lieb loop model. Starting once more from Eq.~(\ref{e:refasm}), we obtain
\begin{multline}
A(n,r)= \sum_{\substack{\varepsilon_1, \ldots, \varepsilon_{n-1} \in \{0,1\}\\
  \sum_{\ell=1}^{n-1} \varepsilon_\ell=r-1}}
\psi^{(n-1)}_{1 + \varepsilon_1, 3 + \varepsilon_2, \ldots,
  2n-3 + \varepsilon_{n-1}} \\
=\sum_{\substack{\varepsilon_1, \ldots, \varepsilon_{n-1} \in \{0,1\}\\
  \sum_{\ell=1}^{n-1} \varepsilon_\ell=r-1}}
\overline{\psi}^{(n-1)}_{2- \varepsilon_1, 4- \varepsilon_2, \ldots,
  2n-2 - \varepsilon_{n-1},2n-1}
=\sum_{\pi:\, \#\{\mathrm{even\ openings\ of\ }\pi\}=r-1} \xi_\pi
\label{e:loopinter}
\end{multline}
as conjectured in the paper \cite{DiFZin07}. In the special case $r=1$ or $r=n$,
the sum in the last right hand side of Eq.~(\ref{e:loopinter})
reduces to a single term, corresponding to one of the two largest
components, with link pattern of the type of fourth or fifth link pattern in Figure~\ref{f:2},
while the left hand side is
simply $A(n,1)=A(n,n)=A(n-1)$. We have thus proved by the same token one of
the conjectures of the paper \cite{BatdeGNie01}. Note also that by summing
Eq.~(\ref{e:loopinter}) over $r$, one obtains on the left hand side the number $A(n)$,
and on the right hand side the sum over all components of the loop model, thus
recovering the main result of the paper \cite{DiFZin05a}.

We are now in a position to prove the third Ttheorem of Section \ref{s:3.3}. In all this section we assume that $\tau=1$.
Let us once again consider the component
$\psi_{1,3,\ldots,2k-3,2k,2k+1,\ldots,2n-1}$, with one defect $2k$ in the sequence of odd indices,
for the XXZ spin chain in size $N=2n+1$. First we can use rotational invariance
to decrease all indices by $2k$ (modulo $N$) and rewrite it as $\psi_{1,3,\ldots,2(n-k)-1,2(n-k+1),\ldots,2n-2,2n+1}$,
that is as a sequence of $n-k$ odd indices, of $k-1$ even indices and $2n+1$ last.
Then we can apply the recurrence relation (\ref{e:rr}):
\begin{multline*}
\psi^{(n)}_{1,3,\ldots,2(n-k)-1,2(n-k+1),\ldots,2n-2,2n+1} \\=
\sum_{\varepsilon_2, \ldots, \varepsilon_n \in \{0,1\}}
\psi^{(n-1)}_{2 -\varepsilon_2,\ldots, 2(n-k)-2-\varepsilon_{n-k},2(n-k)+1-\varepsilon_{n-k+1}, \ldots,2n-3-\varepsilon_{n-1},
2n - \varepsilon_n}.
\end{multline*}
Note that only $\varepsilon_n=1$ contributes to the sum, because the value $2n$ cannot appear in size $2n-1$, the corresponding integral being zero. At this stage one can use the rotational invariance in size $2n-1$ to shift the indices again, this time increasing them by $2k$. The formula we get is:
\[
\psi^{(n)}_{1,3,\ldots,2k-3,2k,2k+1,\ldots,2n-1}=
\sum_{\varepsilon_2, \ldots, \varepsilon_{n-1} \in \{0,1\}}
\psi^{(n-1)}_{2-\varepsilon_{n-k+1}, \ldots,2k-2-\varepsilon_{n-1},2k,2k+2-\varepsilon_2,\ldots,2n-2-\varepsilon_{n-k}}.
\]
In order to identify the right hand side in terms of loops, we must use the complementary set of indices. We find, after reindexation of the summation variables:
\[
\psi^{(n)}_{1,3,\ldots,2k-3,2k,2k+1,\ldots,2n-1}= \sum_{\varepsilon_1, \ldots, \varepsilon_{n-2} \in \{0,1\}} \overline{\psi}^{(n-1)}_{2-\varepsilon_1, \ldots, 2k-2-\varepsilon_{k-1}, 2k-1, 2k+2-\varepsilon_k, \ldots, 2n-2-\varepsilon_{n-2}, 2n-1}.
\]
According to the correspondence described above, see Eq.~(\ref{e:parloop}), the right hand side is nothing but the sum of the components of the Temperley--Lieb loop model eigenvector for which the corresponding link pattern has a closing at $2k$. Hence, the final result is
\[
\psi_{1, 3, \ldots, 2k-3, 2k, 2k+1, \ldots, 2n-1} = \sum_{\pi:\, \pi(2k)<2k} \xi_\pi.
\]
This is in fact the statement of Theorem \ref{c:3}. Indeed, via the reflection that sends $2k$ to~$1$, a closing arch at $2k$ becomes an arch starting at~$1$ and ending somewhere between~$2$ and $2k$.

\subsection{Application to the refined enumeration of ASMs and TSSCPPs}

The main result of the paper \cite{DiFZin07} is the expression of the sum of components of the loop model as a refined enumeration of Totally Symmetric Self-Complementary Plane Partitions (TSSCPPs). More precisely, it reads:
\[
\sum_{\pi:\, \#\{\mathrm{even\ openings\ of\ }\pi\}=r} \xi_\pi =\#\{\mathrm{TSSCPPs}: \#\{i:r_i\ne i \mod 2\}=r\},
\]
where $r_i$ is the endpoint of the $i^{\mathrm th}$ path in the Non-Intersecting Lattice Path formulation of TSSCPPs (see the paper \cite{DiFZin07} for details).

Here we have a similar expression involving a refined enumeration of ASMs, namely Eq.~(\ref{e:loopinter}). Combining the two
we find:
\[
A(n,r)=\#\{\mathrm{TSSCPPs}: \# \{i: r_i \ne i \mod 2\}=r-1\}.
\]
This equality of refined enumerations of ASMs and TSSCPPs was conjectured in \cite{MilRobRum86} (see also \cite{Rob91}: this is the first type of refinement among the three presented).

\section{Conclusion}

We believe that the main result of the present paper are the integral
formulae (\ref{e:b}) and (\ref{e:a}). As we proved at $q = \rme^{\pm 2
\pi \rmi/3}$ they give components of the ground state of the
transfer matrix of the inhomogeneous six-vertex model in the case of
toroidal boundary for an odd number of vertical rows of the
lattice. The homogeneous limit (\ref{e:hl1}) and (\ref{e:hl2}) of the
formulae (\ref{e:b}) and (\ref{e:a}) gives the components of the
ground state vector of the antiferromagnetic XXZ spin chain for the
anisotropy parameter $\Delta$ equal to $-1/2$. This fact has allowed
us to prove some conjectures for these components that were formulated earlier.

On the basis of the coincidence of the expression (\ref{e:a}) with the corresponding homogeneous limit of the integrals considered in the paper \cite{DiFZin07} in the context of Temperley--Lieb loop models, we discovered a connection between the ground state components of odd-sized XXZ spin chains and even-sized Temperley--Lieb loop models. We formulated a new statement relating these components and proved it. We noted that combining the various proved properties of the ground states of XXZ spin chain and loop models lead to interesting results for enumerative combinatorics, in particular a (rather indirect) proof of the
equality of the refined enumeration of ASMs and TSSCPPs.

\subsection*{Acknowledgments}

The work of A.V.R.\ and Yu.G.S.\ was supported in part by the Russian Foundation for Basic Research under grant \# 07--01--00234. P.Z.-J.\ acknowledges the support
of European Marie Curie Research Training Networks ``ENIGMA'' MRT-CT-2004-5652,
``ENRAGE'' MRTN-CT-2004-005616, ESF program ``MISGAM''
and of ANR program ``GIMP'' ANR-05-BLAN-0029-01.

\end{document}